\documentclass{article}\usepackage{pgf,pgfplots}\usetikzlibrary{positioning}
\usepackage{algpseudocode}
\usepackage{algorithm} 
\usepackage{graphicx}
\usepackage{amsmath}
\usepackage{amsfonts}
\usepackage{amssymb}
\usepackage{amsthm}
\usepackage{mathtools}
\usepackage{enumerate}
\usepackage{lscape}
\usepackage{longtable}
\usepackage{rotating}
\usepackage{multirow}
\usepackage{color}
\usepackage{url}
\usepackage{subfigure}
\usepackage{rotating}
\usepackage{comment}
\usepackage[utf8]{inputenc}
\usepackage{amssymb}
\usepackage{amsfonts}
\usepackage{tikz}
\usepackage{natbib}
\usepackage{enumitem}
\usetikzlibrary{arrows.meta, shapes, positioning, fit, backgrounds}

\newtheorem{theorem}{Theorem}

\title{A Survey of Multi Agent Reinforcement Learning \\ Federated Learning and Cooperative and Noncooperative Decentralized Regimes}
\author{Kemboi Cheruiyot, Nickson Kiprotich, Vyacheslav Kungurtsev, \\ Kennedy Mugo, Vivian Mwirigi, Marvin Ngesa }
\date{July 2025}

\begin{document}

\maketitle

\begin{abstract}
    The increasing interest in research and innovation towards the development of autonomous agents presents a number of complex yet important scenarios of multiple AI Agents interacting with each other in an environment. The particular setting can be understood as exhibiting three possibly topologies of interaction - centrally coordinated cooperation, ad-hoc interaction and cooperation, and settings with noncooperative incentive structures. This article presents a comprehensive survey of all three domains, defined under the formalism of Federal Reinforcement Learning (RL), Decentralized RL, and Noncooperative RL, respectively. Highlighting the structural similarities and distinctions, we review the state of the art in these subjects, primarily explored and developed only recently in the literature. We include the formulations as well as known theoretical guarantees and highlights and limitations of numerical performance.
\footnote{Corresponding author \url{vyacheslav.kungurtsev@fel.cvut.cz}. This work acknowledges funding support to the Prosocial AI STEM initiative through the Institute for Advanced Consciousness Studies}\end{abstract}

\section{Introduction}
A process of significant contemporary interest is multiple autonomous AI systems in the same environment as described by Multi-Agent Reinforcement Learning (MARL). There are three distinct circumstances that define particular regimes of interaction: 1) a collection of agents with a means of centralizing communication that are learning about the environment together, defined as \textbf{Federated RL}, 2) a swarm of agents with a network structure of gossip peer-to-peer communication cooperating towards common and individual goals, defined as \textbf{Decentralized MARL} and 3) interacting AI agents within the context of a noncooperative game, which we denote as \textbf{Noncooperative MARL}. 

This Survey presents the state of the art in this field that has largely been developing over, approximately, the last five years. As General Purpose AI become increasingly more capable, Agents engage with an increasing number of administrative computer tasks, and robots and drones become operational in a wider breadth of physical circumstances, understanding the training for cooperation and strategic decision making becomes increasingly critical for ensuring their operation is well-aligned to be safe, effective and robust.

In the broader context of AI Alignment, the additional moving parts of multiple agents presents a greater challenge to establishing forward and backward alignment of user and stakeholder intentions. In addition, potential conflict among two AI Systems corresponding to distinct users can result in unpredictable problems and consequences. Understanding these systems and developing reliable training and communication algorithms will be an active area of research to accommodate the developing multi-agent AI system industry.

We proceed, in the next Section, to provide a brief overview of the main concepts of Reinforcement Learning, We then proceed with our survey. For each class of architecture, we present their form of deployment and structure of communication, the Markov Decision Process (MDP) nomenclature definition of the setting, present formal algorithms representative of the literature, discuss theoretical guarantees, and describe open problems in active or future pursuit by the respective research community.

\begin{itemize}
    \item \textbf{Federated Reinforcement Learning (FRL)}: This constitutes distributed computing client teamwork in solving RL problems, where the clients can reside on distinct autonomous systems. Typically, a central server relays communication between clients and coordinates their training, although all-to-all communication as centralization is also a possible architecture. This involves multiple agents learning locally, keeping data decentralized to preserve privacy of the client. Parameters are exchanged with the eerver after sufficient epochs of local training and agglomerated with some averaging operator by the server, which then relays the centralized parameters back across the clients for the next round of training.
    \item \textbf{Decentralized Multi-Agent Reinforcement Learning (DMARL)}: Here, agents cooperate without a central server, communicating directly peer-to-peer. The structure of available communication channels is defined by a graph, which can be directed or undirected.  This is also known as a Gossip network. They cooperate by communicating parameters in a manner so as to encourage diffusion across the network, while aiming to maximize common and/or individual rewards.
\item \textbf{Noncooperative Multi-Agent Reinforcement Learning (NMARL)}: Alternatively called \emph{Nash MARL}, this defines a set of coupled RL problems wherein the reward function for each agent is distinct and depends on the actions of all of the agents. Typically, there are competitive tradeoffs in the reward structure of the states, with few relevant states that are universally high, and few that are universally low. Understanding and computing various forms of Nash equilibria are described from the literature. 
\end{itemize}

\section{Introduction to Reinforcement Learning}

Reinforcement Learning (RL) is a branch of machine learning that focuses on sequential decision-making, where an agent interacts with an environment to learn an optimal policy through trial and error. Unlike supervised learning, which relies on labeled data, RL involves learning from delayed feedback, where the agent receives rewards or penalties based on its actions, see \cite{sutton2018reinforcement} for a classic text. A key challenge in RL is balancing exploration—where the agent tries new actions to discover better strategies—with exploitation, where it leverages known information to maximize immediate rewards. Striking the right balance between these two aspects is crucial for long-term success in uncertain environments .

Over the decades, RL has evolved from early studies in behavioral psychology and trial-and-error learning to form the backbone of modern algorithms that have achieved remarkable success across diverse applications. High-profile demonstrations, such as the triumph of AlphaGo in defeating world champions and significant advancements in robotics and autonomous driving, highlight the potential of RL to transform complex, real-world decision-making problems \cite{gentle_intro_RL}.

At its core, RL is built upon key concepts such as states, actions, rewards, and policies, underpinned by the mathematical framework of Markov Decision Processes (MDPs). This theoretical foundation not only enables the development of robust algorithms but also highlights inherent challenges, such as sample efficiency, scalability, and the delicate balance between exploration and exploitation.

Recent advances, particularly in deep reinforcement learning, have further expanded the capabilities of RL by integrating deep neural networks with traditional RL methods. This fusion has opened up new avenues for solving problems in high-dimensional spaces and complex environments, while ongoing research continues to address issues like safety, stability, and multi-agent interactions \cite{Rl_and_NNN}.

This section provides a comprehensive overview of Reinforcement Learning, the fundamentals, key algorithms and methods.


\subsection{Markov Decision Processes}
The formal foundation of RL is established through the Markov Decision Process (MDP), a mathematical model for decision-making under uncertainty. An MDP describes an environment where an agent occupies a state, selects an action, transitions to a new state, and receives a corresponding reward. The agent's objective is to determine an optimal policy that maximizes the expected cumulative discounted reward, ensuring that both immediate and future consequences of actions are considered.
The framework of MDPs is central to RL, as it provides a structured way of modeling decision-making problems. An MDP is formally defined as a tuple \( (S, A, P, R, \gamma) \), where:
\begin{itemize}
    \item \( S \) is the set of states,
    \item \( A \) is the set of actions available to the agent,
    \item \( P(s'|s, a) \) is the transition probability from state \( s \) to state \( s' \) given action \( a \),
    \item \( R(s, a) \) is the reward function that provides feedback to the agent,
    \item \( \gamma \in [0,1] \) is the discount factor, which determines the importance of future rewards.
\end{itemize}

The dynamics of MDPs are captured by the transition probabilities \( P(s'|s,a) \), which define the likelihood of transitioning from one state to another given a particular action. The reward function \( R(s, a) \) specifies the immediate benefit of taking an action in a given state, while the discount factor \( \gamma \) balances the trade-off between immediate and future rewards.

A key problem in RL is to find an \textit{optimal policy} \( \pi^* \) that prescribes the best action for each state to maximize the expected cumulative reward. The value function \( V^\pi(s) \) measures the expected return starting from state \( s \) under policy \( \pi \) and is given by:
\[
V^\pi(s) = \sum_{a \in A} \pi(a|s) \left[ R(s, a) + \gamma \sum_{s'} P(s'|s, a) V^\pi(s') \right].
\]
This formulation clearly shows the expectation over both the action selection and the state transition.

\subsection{Dynamic Programming and Bellman Equations}
Dynamic Programming (DP) algorithms solve MDPs by decomposing a complex problem into simpler sub-problems. The Bellman equations are at the heart of DP methods, providing a recursive relationship that expresses the value of a state (or action) in terms of the values of subsequent states.

\paragraph{Value Iteration}
Value Iteration computes the optimal value function \(V^*(s)\) by iteratively applying the Bellman optimality equation:
\[
V^*(s) = \max_{a \in A} \left[ R(s, a) + \gamma \sum_{s'} P(s'|s, a) V^*(s') \right].
\]
This equation indicates that the value of a state is the maximum expected return achievable by considering all possible actions and future states.

\paragraph{Policy Iteration}
Policy Iteration involves two main steps: policy evaluation and policy improvement.

\textbf{Policy Evaluation:}
\[
V^\pi(s) = \sum_{a \in A} \pi(a|s) \left[ R(s, a) + \gamma \sum_{s'} P(s'|s, a) V^\pi(s') \right].
\]

\textbf{Policy Improvement:}
\[
\pi'(s) = \arg\max_{a \in A} \left[ R(s, a) + \gamma \sum_{s'} P(s'|s, a) V^\pi(s') \right].
\]

Although DP methods guarantee convergence, they require complete knowledge of the environment (i.e., the transition probabilities \(P(s'|s,a)\)), which is often impractical in real-world applications.

\subsection{Model-Based and Model-Free Approaches}
Reinforcement learning methods are broadly categorized based on whether they utilize an explicit model of the environment. Model-based methods incorporate or learn a representation of the environment's dynamics, enabling planning via simulated trajectories. This approach often results in superior sample efficiency since the agent can extract additional learning from simulated experiences. However, the success of model-based methods hinges on the accuracy of the learned model, and inaccuracies may lead to suboptimal decisions.

In contrast, model-free methods learn value functions or policies directly from raw experiences, avoiding the need to model transition dynamics explicitly. Techniques such as Monte Carlo methods and Temporal Difference (TD) learning exemplify this approach. While model-free methods are generally simpler and more robust in complex or stochastic environments, they typically require more interactions with the environment to converge to an optimal solution. Recent developments—such as Dyna architectures—seek to combine these two paradigms, leveraging the sample efficiency of model-based approaches and the robustness of model-free methods.

\subsection{Temporal Difference Learning Methods}
Temporal difference (TD) learning stands out as a pivotal algorithmic breakthrough in reinforcement learning, merging ideas from dynamic programming and Monte Carlo methods. TD methods update value estimates based on the immediate reward and the estimated value of subsequent states—a process known as bootstrapping. For example, the TD(0) update rule is given by:
\[
V(s_t) \leftarrow V(s_t) + \alpha \Bigl[r_{t+1} + \gamma V(s_{t+1}) - V(s_t)\Bigr],
\]
where \(\alpha\) is the learning rate and the term in brackets is the TD error.

For control problems, TD learning has been extended to methods such as Sarsa (an on-policy method) and Q-learning (an off-policy method). Sarsa updates its action-value function as:
\[
Q(s_t, a_t) \leftarrow Q(s_t, a_t) + \alpha\Bigl[r_{t+1} + \gamma Q(s_{t+1}, a_{t+1}) - Q(s_t, a_t)\Bigr],
\]
while Q-learning uses the update:
\[
Q(s_t, a_t) \leftarrow Q(s_t, a_t) + \alpha\Bigl[r_{t+1} + \gamma \max_a Q(s_{t+1}, a) - Q(s_t, a_t)\Bigr].
\]
Both approaches have been shown to converge under suitable conditions, with Q-learning providing the flexibility to learn about optimal policies while following exploratory strategies.

TD learning further extends to TD(\(\lambda\)) methods by introducing eligibility traces, which bridge the gap between one-step TD and Monte Carlo methods. The forward view of TD(\(\lambda\)) defines the \(\lambda\)-return as:
\[
G_t^\lambda = (1-\lambda) \sum_{n=1}^\infty \lambda^{n-1} G_t^{(n)},
\]
with the n-step return
\[
G_t^{(n)} = \sum_{k=0}^{n-1} \gamma^k r_{t+k} + \gamma^n V(s_{t+n}).
\]
The backward view implements this idea efficiently via eligibility traces, updating the value function as:
\[
V(s) \leftarrow V(s) + \alpha\, \delta_t\, e(s),
\]
where the eligibility trace \(e(s)\) is updated by:
\[
e(s) \leftarrow \gamma \lambda \, e(s) + \mathbf{1}\{s_t = s\}.
\]
This approach has been instrumental in achieving improved convergence properties and handling both episodic and continuing tasks \cite{sutton2018reinforcement,arxiv2312}.

\subsection{Policy Optimization and Actor-Critic Methods}
While value-based methods focus on estimating state or state-action values, policy-based methods directly optimize the policy. In Policy Gradient methods, the agent learns a parameterized policy and adjusts the parameters according to the gradient of the expected reward. The policy gradient theorem provides the foundation:
\[
\nabla_\theta J(\theta) = \mathbb{E}_\pi \left[ \nabla_\theta \log \pi_\theta(a|s) \, Q^\pi(s,a) \right].
\]
A common variant REINFORCE introduces a baseline \( b(s) \) to reduce variance:
\[
\nabla_\theta J(\theta) = \mathbb{E}_\pi \left[ \nabla_\theta \log \pi_\theta(a|s) \left( Q^\pi(s,a) - b(s) \right) \right].
\]

Policy gradient methods are appealing because they directly optimize the policy without the need for an explicit value function. The reinforce algorithm is a Monte Carlo approach that uses the following update:
\[
\theta_{t+1} = \theta_t + \alpha \, \nabla_\theta \log \pi_\theta(a_t|s_t) \, G_t,
\]
where \( G_t \) is the return following time \( t \).

Variance reduction is critical in practice. One effective approach is the use of an \emph{advantage function} \( A^\pi(s,a) = Q^\pi(s,a) - V^\pi(s) \), which measures how much better an action is compared to the average. This leads to the following update:
\[
\theta_{t+1} = \theta_t + \alpha \, \nabla_\theta \log \pi_\theta(a_t|s_t) \, A^\pi(s_t,a_t).
\]
Actor-Critic methods combine these ideas by using a critic to estimate \( V^\pi(s) \) or \( Q^\pi(s,a) \) and an actor to update the policy. Recent advances in this area have led to algorithms with improved convergence properties and sample efficiency \citep{arxiv2109, arxiv2011}.

\subsection{Epsilon-Greedy Policies and the Exploration-Exploitation Tradeoff}
A fundamental challenge in reinforcement learning is managing the trade-off between exploration (seeking new information) and exploitation (using known information). A common strategy, the \(\epsilon\)-greedy policy, selects the best-known action with probability \(1-\epsilon\) while choosing a random action with probability \(\epsilon\):
\[
\pi(a|s) =
\begin{cases}
1 - \epsilon + \frac{\epsilon}{|A|}, & \text{if } a = \arg\max_{a'} Q(s, a'), \\
\frac{\epsilon}{|A|}, & \text{otherwise}.
\end{cases}
\]
Beyond \(\epsilon\)-greedy, more sophisticated exploration techniques—such as softmax action selection, optimistic initialization, Upper Confidence Bounds (UCB), Thompson sampling, and intrinsic motivation strategies—have been developed. These methods address challenges like sparse rewards, deceptive local optima, and long-horizon planning, and they are crucial in complex environments where efficient exploration is essential for learning robust policies\cite{sutton2018reinforcement}.

\subsection{Stochastic Approximation and Two-Time Scale Methods}
Stochastic Approximation is a class of iterative methods for solving optimization problems when the underlying function is noisy. In RL, these methods are particularly useful when the environment’s model is unknown or dynamic. Two-time scale methods enhance stochastic approximation by employing different learning rates for the policy and value function updates, which can lead to improved stability and efficiency.
A general stochastic approximation update is given by:
\[
\theta_{t+1} = \theta_t + \alpha_t \Bigl( h(\theta_t) + M_{t+1} \Bigr),
\]
where \( \alpha_t \) is a diminishing step-size, \( h(\theta_t) \) is the mean update, and \( M_{t+1} \) represents noise.

In many RL algorithms (notably Actor-Critic), different parameters are updated at different rates. Suppose we have two sets of parameters: \( \theta \) (e.g., for the actor) and \( \omega \) (e.g., for the critic). Their updates are performed on different time scales:
\begin{align}
  \theta_{t+1} &= \theta_t + \alpha_t \Bigl( h(\theta_t, \omega_t) + M_{t+1}^{\theta} \Bigr), \label{eq:two_time_scale_actor_exp} \\[1mm]
  \omega_{t+1} &= \omega_t + \beta_t \Bigl( g(\theta_t, \omega_t) + M_{t+1}^{\omega} \Bigr), \label{eq:two_time_scale_critic_exp}
\end{align}
with the step-sizes satisfying
\[
\lim_{t \to \infty} \frac{\alpha_t}{\beta_t} = 0.
\]
This condition ensures that \( \omega_t \) (the critic) converges much faster than \( \theta_t \) (the actor), so that the actor’s updates effectively see a near-converged critic. Two-time scale stochastic approximation has been rigorously analyzed using ordinary differential equation (ODE) techniques, and its general convergence properties have been established in recent works \cite{doan2019, siam22, researchgate_two_time}.

These techniques are essential in ensuring the stability of complex RL algorithms where multiple interacting components are updated simultaneously, particularly in non-stationary or high-dimensional settings.

A major theoretical result in this area is provided by Borkar's Two Time-Scale Convergence Theorem. One common version of the theorem is stated as follows:

\begin{theorem}[Borkar's Two Time-Scale Convergence Theorem \cite{borkar2008stochastic}]
Consider the two-time scale stochastic approximation iterates given by
\begin{align*}
\theta_{t+1} &= \theta_t + \alpha_t \left( h(\theta_t, \omega_t) + M_{t+1}^\theta \right), \\
\omega_{t+1} &= \omega_t + \beta_t \left( g(\theta_t, \omega_t) + M_{t+1}^\omega \right),
\end{align*}
where the step-sizes \(\{\alpha_t\}\) and \(\{\beta_t\}\) are positive, diminishing, and satisfy
\[
\lim_{t\to\infty} \frac{\alpha_t}{\beta_t} = 0.
\]
Under standard regularity conditions (including Lipschitz continuity of \(h\) and \(g\), boundedness of the iterates, and appropriate martingale difference noise conditions), the following holds:
\begin{enumerate}
    \item The fast iterate \(\omega_t\) converges to a unique stable equilibrium \(\omega^*(\theta)\) of the ODE
    \[
    \dot{\omega}(t) = g(\theta, \omega(t))
    \]
    for a fixed \(\theta\).
    \item The slow iterate \(\theta_t\) converges to an invariant set of the ODE
    \[
    \dot{\theta}(t) = h\bigl(\theta(t), \omega^*(\theta(t))\bigr).
    \]
\end{enumerate}
\end{theorem}
This theorem is pivotal in the analysis of Actor-Critic algorithms and other multi-time scale methods, ensuring that the slower parameter updates see a quasi-static environment as the faster updates converge rapidly. Such convergence guarantees are crucial in establishing the reliability and stability of learning algorithms in dynamic environments.

\subsection{Summary}
Reinforcement Learning is a dynamic and rapidly evolving field built upon the formalism of Markov Decision Processes, Dynamic Programming, and Temporal Difference learning. The integration of model-free methods, action-value approaches, and policy optimization techniques (including actor-critic and policy gradient methods) has enabled RL to tackle complex, high-dimensional decision-making tasks. Moreover, advanced tools such as TD(\(\lambda\)) for efficient credit assignment and two-time scale stochastic approximation for stability in multi-parameter updates have been instrumental in the development of robust RL algorithms. As RL continues to find applications in robotics, game playing, healthcare, and autonomous systems, ongoing research remains focused on improving sample efficiency, convergence properties, and scalability.

\section{Federated RL (FRL)}

\subsection{Introduction to FRL}

Federated Reinforcement Learning (FRL) emerges as an innovative approach that integrates the privacy-preserving mechanisms of Federated Learning (FL) with the decision-making capabilities of Reinforcement Learning (RL) \cite{qi2021federated}. It represents a significant advancement in machine learning that addresses critical challenges in modern distributed systems where data privacy is paramount. According to the paper, FRL can be defined as "an integration of FL and RL under privacy protection, where several elements of RL can be presented in FL frameworks to deals with sequential decision-making tasks" \cite{qi2021federated}.

The fundamental architecture of FRL maintains the core principles of federated systems where multiple agents collaborate to build a shared model without directly exchanging their private data \cite{qi2021federated}. Instead, these agents share model parameters or gradients, allowing the system to leverage collective experiences while maintaining data locality and privacy. This collaborative approach enables agents to benefit from the experiences of others without compromising sensitive information.

FRL algorithms are categorized into two main types based on the distribution characteristics of agents within the framework: Horizontal Federated Reinforcement Learning (HFRL) and Vertical Federated Reinforcement Learning (VFRL) \cite{qi2021federated}. This categorization mirrors the approach used in traditional federated learning, where similar distinctions are made based on data distribution patterns. The paper elaborates on these categories by providing specific case studies, including applications in autonomous driving and smart grid systems, to illustrate the practical implementation differences between HFRL and VFRL approaches~\cite{qi2021federated}.

\subsubsection{Core Components of FRL Systems}

The FRL framework consists of distributed agents operating in potentially different environments, a coordination mechanism for model aggregation, and a secure communication protocol \cite{qi2021federated}. Each agent maintains its own local model trained on private experience data, while the aggregation process combines these models to create a globally improved policy that benefits all participants. This distributed architecture allows FRL to scale effectively across numerous devices or systems while maintaining robust privacy protections.

\subsubsection{Relationship Between FL and RL}
The integration of FL and RL creates a synergistic relationship that addresses limitations in both fields. FL contributes its distributed learning framework and privacy-preserving mechanisms, while RL provides methods for sequential decision-making in dynamic environments through interaction and feedback \cite{qi2021federated}.

Federated Learning emerged as a solution to privacy concerns in traditional machine learning approaches. It enables multiple parties to collaboratively train a model without sharing their raw data \cite{qi2021federated}. As described in the paper, FL is "a decentralized collaborative approach that allows multiple partners to train data respectively and build a shared model while maintaining privacy" \cite{qi2021federated}. The FL process typically involves local model training, secure aggregation of model updates, and redistribution of the improved model to participants.

Reinforcement Learning, conversely, focuses on how agents learn optimal behaviors through interactions with an environment. The paper defines RL as "one of the branches [of machine learning] that focuses on how individuals, i.e., agents, interact with their environment and maximize some portion of the cumulative reward" \cite{qi2021federated}. This learning process occurs through trial and error, with the agent developing policies that maximize expected rewards over time.

\subsubsection{Integrating Reinforcement Learning Principles with Federated Architectures}

Building upon the foundational concepts of Reinforcement Learning previously established, we now examine how these principles extend to federated environments. The integration of RL mechanisms with federated architectures necessitates reconceptualizing several core aspects of the traditional RL framework to accommodate distributed learning while preserving data privacy.

\paragraph{Markov Decision Processes in FRL}
In federated reinforcement learning, the traditional Markov Decision Process (MDP) framework requires adaptation to account for distributed agents operating under data isolation constraints. An FRL system can be formalized as a collection of local MDPs, where each agent $i$ maintains its own tuple $(\mathcal{S}_i, \mathcal{A}_i, \mathcal{P}_i, \mathcal{R}_i, \gamma)$ representing states, actions, transition probabilities, rewards, and discount factor, respectively \cite{qi2021federated}.

The key distinction emerges in how these local MDPs interact within the federated framework.  \cite{mcmahan2017communication}, who introduced the seminal FedAvg algorithm, provide the foundation for parameter aggregation in federated settings. Building upon and extending these foundations, \cite{jin2022fed} propose that policy interactions in FRL can be modeled as:

\[
\pi_{\text{global}} = \mathcal{F}(\{\pi_i\}_{i=1}^N, \{w_i\}_{i=1}^N)
\]
where $\pi_i$ represents the policy of agent $i$, $w_i$ is the weight assigned to agent $i$ (typically proportional to the size of its local dataset), and $\mathcal{F}$ is the federation function that aggregates local policies into a global policy $\pi_{\text{global}}$ using weighted averaging similar to FedAvg.

\paragraph{Value Function Decomposition in FRL}
Following policy aggregation, value function decomposition becomes essential in federated contexts. Rather than maintaining a single centralized value function, FRL systems operate with a distributed set of value functions that must be coherently aggregated. The work \cite{zhuo2019federated} formulates this relationship through a weighted averaging approach:

\[
V^{\pi}_{\text{global}}(s) = \sum_{i=1}^{N} \frac{n_i}{n_{\text{total}}} V^{\pi}_i(s)
\]
where $V^{\pi}_i(s)$ represents the value function of agent $i$ under policy $\pi$ for state $s$, $n_i$ represents the number of samples from agent $i$, and $n_{\text{total}} = \sum_{i=1}^{N} n_i$ is the total number of samples across all agents. This approach ensures that agents with more extensive experience have proportionally greater influence on the global value function, similar to the FedAvg mechanism proposed by \cite{mcmahan2017communication}.

\paragraph{Federated Policy Optimization}
The policy optimization process in FRL requires balancing local improvements with global convergence. Chen et al. \cite{chen2021federated} propose a federated policy gradient approach that modifies the traditional policy gradient theorem to accommodate federated constraints:

\begin{equation}
\nabla_{\theta} J(\pi_{\theta}) = \mathbb{E}_{\pi_{\theta}} \left[ \sum_{i=1}^{N} \alpha_i \nabla_{\theta} \log \pi_{\theta}(a|s) Q^{\pi}_i(s,a) \right]
\end{equation}

where $\alpha_i$ represents the contribution factor of agent $i$ (typically proportional to dataset size), and $Q^{\pi}_i(s,a)$ is the local action-value function. This approach enables agents to optimize policies based on their local experiences while contributing to a globally improved policy. Importantly, Chen et al. provide theoretical convergence guarantees for this method under certain assumptions, although convergence rates may degrade with highly non-IID data distributions across agents.

\paragraph{Privacy-Preserving Experience Replay}
Experience replay, a technique that enhances sample efficiency in RL by reusing past experiences, faces unique challenges in federated settings due to privacy constraints. Traditional experience sharing is prohibited in FRL, necessitating alternative approaches. \cite{truex2019hybrid} introduce hybrid differential privacy mechanisms applicable to federated learning, which \cite{wei2020federated} extend to reinforcement learning contexts:

\begin{equation}
D_{\text{fed}} = \psi(\{D_i\}_{i=1}^N, \epsilon, \delta)
\end{equation}

where $D_i$ represents the local experience buffer of agent $i$, and $\psi$ is a privacy-preserving aggregation function that uses differential privacy with privacy budget parameters $\epsilon$ and $\delta$. This approach enables knowledge transfer without raw data exchange by adding calibrated noise to statistical summaries rather than sharing raw experiences, thereby maintaining $(\epsilon, \delta)$-differential privacy guarantees.

\paragraph{Federated Q-Learning}
Building on experience replay mechanisms, Q-learning adapts to federated environments through modifications to its update rule. \cite{nadiger2019federated} propose a federated Q-learning approach with the following update mechanism:

\begin{equation}
Q_{\text{global}}(s, a) \leftarrow (1 - \beta) Q_{\text{global}}(s, a) + \beta \sum_{i=1}^{N} \frac{n_i}{n_{\text{total}}} Q_i(s, a)
\end{equation}

where $\beta$ is the global learning rate, $n_i$ is the number of samples contributed by agent $i$, and $n_{\text{total}}$ is the total number of samples across all agents. This approach weights the contributions of agents based on their local dataset sizes. Notably, Nadiger et al. acknowledge an important challenge: when agents have conflicting Q-values for the same state-action pairs due to environment differences, simple averaging may lead to suboptimal policies. They propose a confidence-weighted aggregation as a partial solution, but recognize this as an ongoing research challenge in heterogeneous FRL settings.

\subsubsection{Mapping FL and RL Components}

FRL leverages FL's decentralized learning framework while addressing sequential decision-making tasks inherent in RL. The components of FL are mapped to RL elements as follows:
\begin{itemize}
    \item The environment in Reinforcement Learning (RL) corresponds to the sample space in Federated Learning (FL).
    \item States observed by agents correspond to features in traditional FL.
    \item Actions selected by agents correspond to labels or predictions in FL models.
\end{itemize}

This mapping ensures that FRL retains the strengths of both FL and RL while addressing their limitations \cite{qi2021federated}.

By integrating these RL principles into FL frameworks, FRL enables distributed agents to collaboratively solve complex sequential decision-making tasks while preserving privacy. This synergy between FL and RL represents a significant advancement in machine learning for modern distributed systems.

\subsection{FRL Architecture and Categories}

According to the distribution characteristics of agents within the framework, FRL architectures can be systematically classified into two fundamental categories:

\begin{enumerate}
    \item \textbf{Horizontal Federated Reinforcement Learning (HFRL)}: Addresses scenarios where agents share the same state-action spaces but possess different experiences or data distributions.
    
    \item \textbf{Vertical Federated Reinforcement Learning (VFRL)}: Manages scenarios where different agents observe different features or aspects of the environment, requiring integration of partial observations.
\end{enumerate}

This categorization parallels the classification in Federated Learning between Horizontal Federated Learning (HFL) and Vertical Federated Learning (VFL), but adapts to the sequential decision-making context of reinforcement learning.

\subsubsection{Horizontal Federated Reinforcement Learning (HFRL)}

\paragraph{Theoretical Foundation} 
HFRL addresses scenarios where multiple agents operate in similar environments with consistent state-action spaces but different experiences or data distributions. The autonomous driving case exemplifies this architecture, where vehicles navigate similar action and state spaces but encounter different environmental conditions.

In HFRL, the agents share the same state and action spaces, but may experience different transition dynamics and reward functions based on their unique environment interactions:

\begin{equation}
\forall i, j: S_i = S_j \text{ and } A_i = A_j
\end{equation}

But potentially different transition and reward functions:

\begin{equation}
\exists i, j: P_i \neq P_j \text{ or } R_i \neq R_j
\end{equation}

\paragraph{Mathematical Formulation}

For an HFRL system with $N$ agents, each agent $i$ operates within its own MDP:

\begin{equation}
\text{MDP}_i = (S, A, P_i, R_i, \gamma)
\end{equation}

Each agent aims to maximize its expected return based on a policy $\pi_i$:

\begin{equation}
J(\pi_i) = \mathbb{E}_{\tau \sim \pi_i}\left[\sum_{t=0}^{\infty} \gamma^t R_i(s_t, a_t, s_{t+1})\right]
\end{equation}

Where $\tau$ represents trajectories generated by following policy $\pi_i$.

\paragraph{Value Function Aggregation}

Using Q-learning as an example implementation, each agent maintains a local Q-function $Q_i(s, a)$ that is updated through local experiences:

\begin{equation}
Q_i(s_t, a_t) \leftarrow Q_i(s_t, a_t) + \alpha \left[r_t + \gamma \max_{a} Q_i(s_{t+1}, a) - Q_i(s_t, a_t)\right]
\end{equation}

The federated aggregation periodically combines these local functions into a global Q-function:

\begin{equation}
Q_{\text{global}}(s, a) = \sum_{i=1}^{N} \frac{n_i}{n} Q_i(s, a)
\end{equation}

This aggregated function is then distributed back to all agents for continued local learning.

\paragraph{Policy-Based Formulation}

For policy-based methods, each agent maintains a parameterized policy $\pi_i(a|s; \theta_i)$ where $\theta_i$ represents policy parameters. The local objective function for agent $i$ is:

\begin{equation}
J_i(\theta_i) = \mathbb{E}_{s \sim \rho^{\pi_i}, a \sim \pi_i(\cdot|s; \theta_i)}\left[Q^{\pi_i}(s, a)\right]
\end{equation}

Where $\rho^{\pi_i}$ denotes the state distribution under policy $\pi_i$. The local policy parameters are updated using gradient ascent:

\begin{equation}
\theta_i \leftarrow \theta_i + \beta \nabla_{\theta_i} J_i(\theta_i)
\end{equation}

The federation process aggregates these parameters:

\begin{equation}
\theta_{\text{global}} = \sum_{i=1}^{N} \frac{n_i}{n} \theta_i
\end{equation}

This approach enables collaborative policy improvement while preserving the privacy of local experiences.

\clearpage

\begin{figure}[h]
    \centering
    \begin{tikzpicture}
        \draw[dashed] (0,0) rectangle (6,2);
        \draw[dashed] (1,-1) rectangle (7,1);
        \draw[dashed] (2,-2) rectangle (8,0);
        \node at (1.2,1.5) {Local Model A, Agent A, Env A};
        \node at (2.2,0.5) {Local Model B, Agent B, Env B};
        \node at (3.2,-0.5) {Local Model C, Agent C, Env C};

    \end{tikzpicture}
    \caption{Horizontal Federated Reinforcement Learning }
\end{figure}
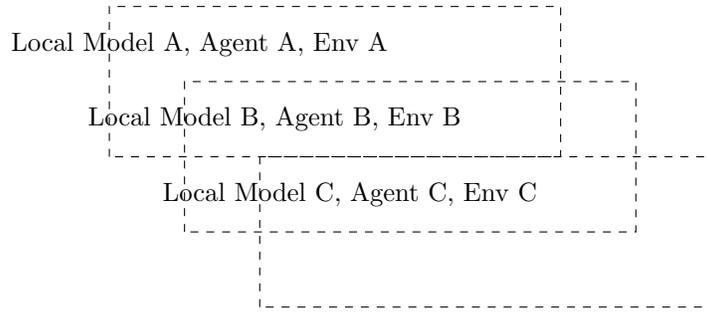

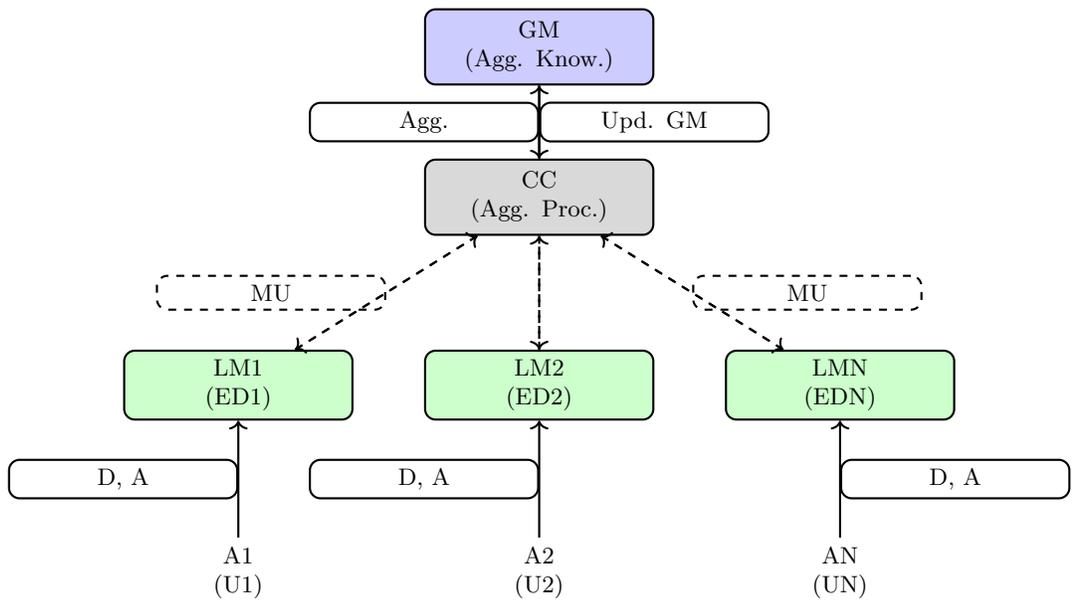
\begin{figure}[h]
    \centering
    \begin{tikzpicture}[
        node distance=1.5cm and 2cm,
        every node/.style={draw, text width=2.8cm, align=center, rounded corners, font=\small},
        every path/.style={thick, ->}
    ]

        \node[fill=blue!20, minimum height=1cm] (GM) at (4,7) {GM \\ (Agg. Know.)};

        \node[fill=gray!30, minimum height=1.0cm] (CC) at (4,5) {CC \\ (Agg. Proc.)};

        \node[fill=green!20, minimum height=0.8cm] (LM1) at (0,2.5) {LM1 \\ (ED1)};
        \node[fill=green!20, minimum height=0.8cm] (LM2) at (4,2.5) {LM2 \\ (ED2)};
        \node[fill=green!20, minimum height=0.8cm] (LMN) at (8,2.5) {LMN \\ (EDN)};

        \node[draw=none] (A1) at (0,0) {A1 \\ (U1)};
        \node[draw=none] (A2) at (4,0) {A2 \\ (U2)};
        \node[draw=none] (AN) at (8,0) {AN \\ (UN)};

        \draw[->] (A1) -- (LM1) node[midway, left] {D, A};
        \draw[->] (A2) -- (LM2) node[midway, left] {D, A};
        \draw[->] (AN) -- (LMN) node[midway, right] {D, A};

        \draw[->, dashed] (LM1) -- (CC) node[midway, left] {MU};
        \draw[->, dashed] (LM2) -- (CC);
        \draw[->, dashed] (LMN) -- (CC) node[midway, right] {MU};

        \draw[->, thick] (CC) -- (GM) node[midway, left] {Agg.};

        \draw[->, thick] (GM) -- (CC) node[midway, right] {Upd. GM};

        \draw[->, dashed] (CC) -- (LM1);
        \draw[->, dashed] (CC) -- (LM2);
        \draw[->, dashed] (CC) -- (LMN);

    \end{tikzpicture}
    \caption{ An example of horizontal federated reinforcement learning (HFRL) architecture.}
\end{figure}

\vspace{0.8cm}
\renewcommand{\arraystretch}{1.3}
\begin{center}
\begin{tabular}{|c|l|} 
    \hline
    \textbf{Abbr.} & \textbf{Meaning} \\ \hline
    GM  & Global Model \\ \hline
    CC  & Central Coordinator \\ \hline
    LM  & Local Model \\ \hline
    ED  & Edge Device \\ \hline
    A   & Agent \\ \hline
    U   & User \\ \hline
    D   & Data \\ \hline
    MU  & Model Update \\ \hline
    Agg. & Aggregation \\ \hline
    Upd. GM & Updated Global Model \\ \hline
    
\end{tabular}
\end{center}

\subsubsection{Vertical Federated Reinforcement Learning (VFRL)}
\paragraph{Theoretical Foundation}

VFRL addresses scenarios where different agents observe different features or aspects of the environment, requiring integration of these partial observations. The smart grid case exemplifies this architecture, where different components of the grid observe different aspects of the overall system state.

In VFRL, the environment state $s$ is partitioned across agents, with each agent $i$ observing only a subset $s^i$ of the complete state:

\begin{equation}
s = \{s^1, s^2, \ldots, s^N\}
\end{equation}

Similarly, the action space may be partitioned:

\begin{equation}
a = \{a^1, a^2, \ldots, a^N\}
\end{equation}

This can be formally characterized as:

\begin{equation}
\forall i \neq j: S^i \cap S^j = \emptyset \text{ or } A^i \cap A^j = \emptyset
\end{equation}

This formulation represents the extreme case of complete partitioning, though partial overlaps may exist in practical implementations.

\paragraph{Mathematical Formulation}

The complete MDP for a VFRL system can be represented as:

\begin{equation}
\text{MDP} = (S = S^1 \times S^2 \times \ldots \times S^N, A = A^1 \times A^2 \times \ldots \times A^N, P, R, \gamma)
\end{equation}

Each agent maintains a local value function or policy based on its partial observation. For value-based methods, agent $i$ maintains $Q_i(s^i, a)$, which estimates expected returns given its observed state component.

\paragraph{Feature-Based Integration}

A sophisticated VFRL approach involves agents exchanging feature representations rather than direct value functions. Each agent $i$ maintains an encoder network $E_i(s^i)$ that maps its partial observation to a feature vector $\phi_i$. These feature vectors are exchanged and combined:

\begin{equation}
\phi = g(\phi_1, \phi_2, \ldots, \phi_N)
\end{equation}

Where $g$ is a feature aggregation function. This combined representation is then used to estimate the value function or policy:

\begin{equation}
Q(s, a) = h(\phi, a)
\end{equation}

The parameters of encoders $E_i$, aggregator $g$, and value estimator $h$ are jointly optimized through federation while preserving privacy of raw observations.

\paragraph{Combined Value Function Approximation}

The federated process in VFRL involves combining partial value functions to approximate the global value function:

\begin{equation}
Q(s, a) \approx f(Q_1(s^1, a), Q_2(s^2, a), \ldots, Q_N(s^N, a))
\end{equation}

Where $f$ represents an aggregation function that combines the partial value estimates. This function could be a simple average, a weighted sum, or a more complex neural network that learns to optimally combine the partial estimates.

\vspace{1cm}

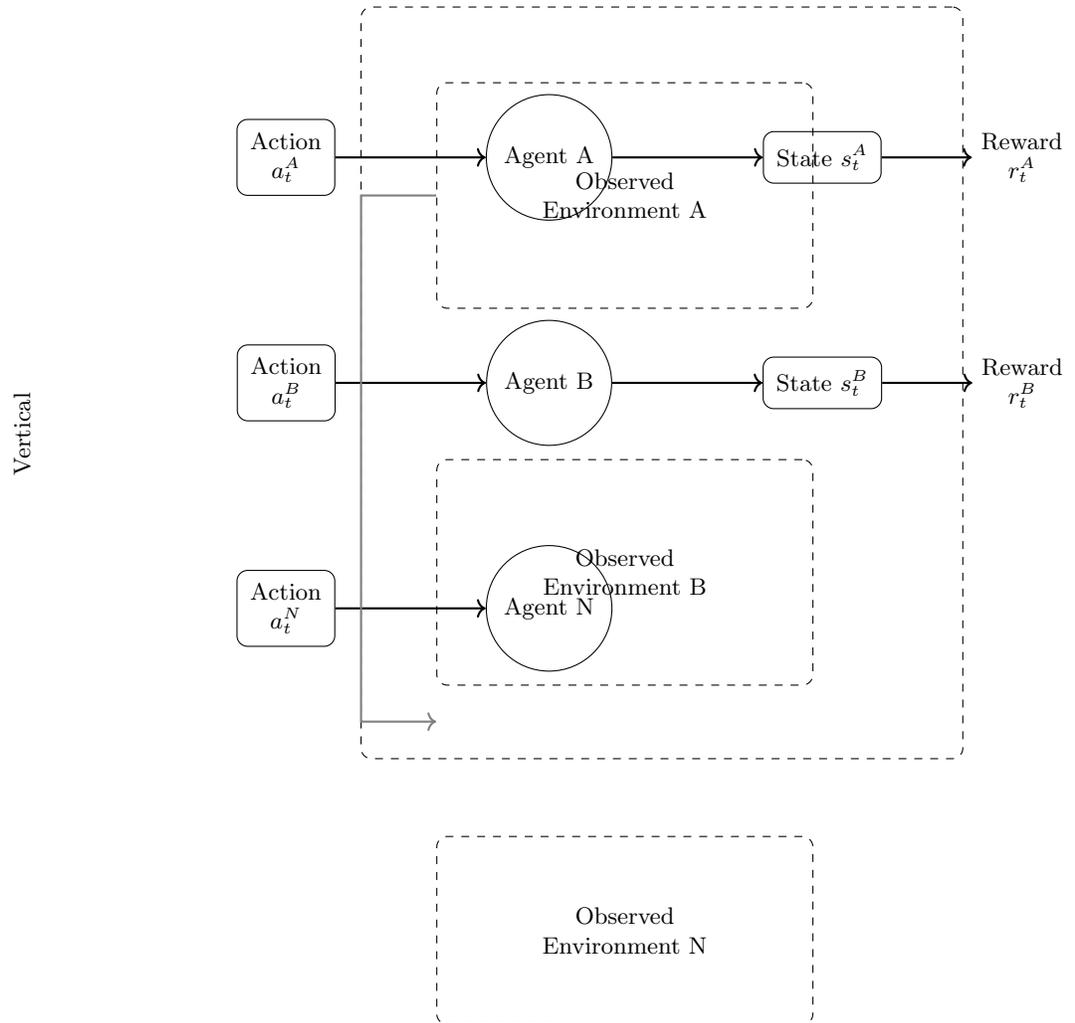
\begin{figure}
    \centering
    \begin{tikzpicture}[
        font=\small,
        every node/.style={align=center},
        box/.style={draw, dashed, rounded corners, inner sep=10pt},
        arrow/.style={->, thick},
        env/.style={draw, rectangle, inner sep=5pt, rounded corners},
        agent/.style={draw, circle, inner sep=5pt},
        global/.style={draw, thick, gray, ->}
    ]
    
    \node[box, minimum width=8cm, minimum height=10cm] (global) at (0,0) {};
    
    \node[box, minimum width=5cm, minimum height=3cm, anchor=north west] (envA) at (-3,4) {Observed \\ Environment A};
    \node[agent] (agentA) at (-1.5,3) {Agent A};
    \node[env, right=2cm of agentA] (stateA) {State $s_t^A$};
    \node[right=1.2cm of stateA] (rewardA) {Reward \\ $r_t^A$};
    
    \draw[arrow] (agentA.east) -- (stateA.west);
    \draw[arrow] (stateA.east) -- (rewardA.west);
    
    \node[env, left=2cm of agentA] (actionA) {Action \\ $a_t^A$};
    \draw[arrow] (actionA.east) -- (agentA.west);
    
    \node[box, minimum width=5cm, minimum height=3cm, below=2cm of envA] (envB) {Observed \\ Environment B};
    \node[agent] (agentB) at (-1.5,0) {Agent B};
    \node[env, right=2cm of agentB] (stateB) {State $s_t^B$};
    \node[right=1.2cm of stateB] (rewardB) {Reward \\ $r_t^B$};
    
    \draw[arrow] (agentB.east) -- (stateB.west);
    \draw[arrow] (stateB.east) -- (rewardB.west);
    
    \node[env, left=2cm of agentB] (actionB) {Action \\ $a_t^B$};
    \draw[arrow] (actionB.east) -- (agentB.west);
    
    \node[box, minimum width=5cm, minimum height=2.5cm, below=2cm of envB] (envN) {Observed \\ Environment N};
    \node[agent] (agentN) at (-1.5,-3) {Agent N};
    
    \node[env, left=2cm of agentN] (actionN) {Action \\ $a_t^N$};
    \draw[arrow] (actionN.east) -- (agentN.west);
    
    \draw[global] (envA.west) -- ++(-1,0) -- ++(0,-7) -- ++(1,0);
    
    \node[rotate=90, left=4.5cm of global] {Vertical};
    
    \end{tikzpicture}
    
    \caption{Vertical Federated Reinforcement Learning}
\end{figure}
\clearpage

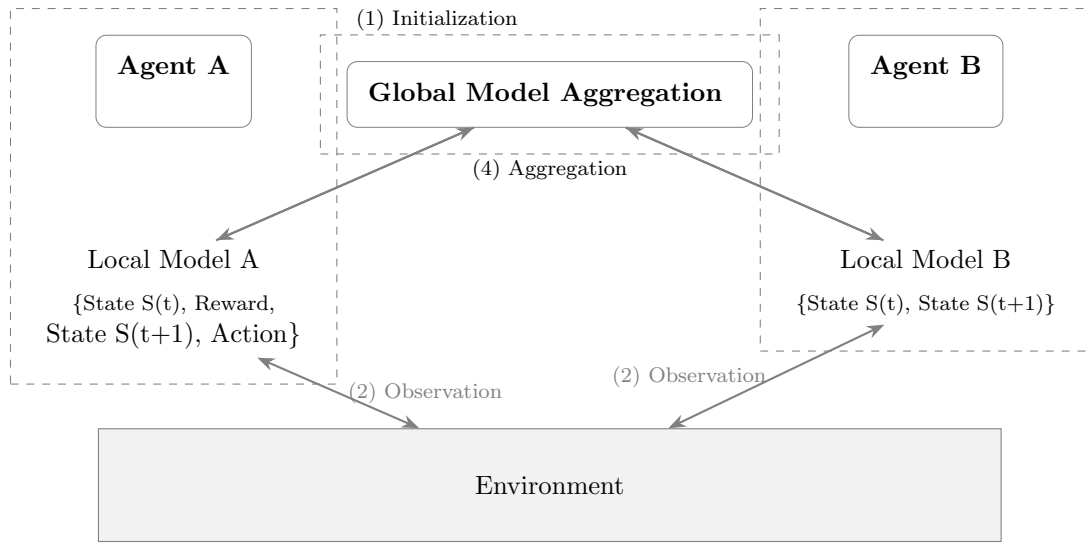
\begin{figure}[h]
    \begin{tikzpicture}[
      >=Stealth,
      every node/.style={align=center},
      box/.style={rectangle, rounded corners, draw=gray, very thin, inner sep=8pt},
      arrow/.style={->, thick, gray}, 
      dashedbox/.style={draw=gray, dashed, inner sep=10pt}
    ]
    
      \node (environment) [rectangle, draw=gray, fill=gray!10, minimum width=12cm, minimum height=1.5cm] {Environment};
    
      \node (agent_a) [box, above=4cm of environment, xshift=-5cm] { 
        \textbf{Agent A} \\
      };
      \node (local_model_a) [below=1.5cm of agent_a] {Local Model A};
      \node (train_data_a) [below=0.1cm of local_model_a] {\footnotesize \{State S(t), Reward, \\ State S(t+1), Action\}};
    
      \node (agent_b) [box, above=4cm of environment, xshift=5cm] { 
        \textbf{Agent B} \\
      };
      \node (local_model_b) [below=1.5cm of agent_b] {Local Model B};
      \node (input_data_b) [below=0.1cm of local_model_b] {\footnotesize \{State S(t), State S(t+1)\}};
    
      \node (global_model) [box, above=of environment, yshift=3cm] { 
        \textbf{Global Model Aggregation}
      };
    
      \begin{scope}[font=\footnotesize]
        \draw[arrow, <->] (environment) -- node[midway, right] {(2) Observation} (train_data_a);
        \draw[arrow] (local_model_a) -- node[midway, above] {} (global_model);
        \draw[arrow] (global_model) -- node[midway, below] {} (local_model_a);
        \node at ([shift={(1.0cm,0.75cm)}]local_model_a) {};

        \draw[arrow, <->] (environment) -- node[midway, left] {(2) Observation} (input_data_b);
        \draw[arrow] (local_model_b) -- node[midway, above] {} (global_model);
        \draw[arrow] (global_model) -- node[midway, below] {} (local_model_b);
        \node at ([shift={(-1.0cm,0.75cm)}]local_model_b) {};
        
        \node at ([shift={(-1.5cm,1.0cm)}]global_model) {(1) Initialization};
        \node at ([shift={(0,-1cm)}]global_model) {(4) Aggregation};
      \end{scope}
    
      \node [dashedbox, fit=(agent_a) (local_model_a) (train_data_a)] {};
      \node [dashedbox, fit=(agent_b) (local_model_b) (input_data_b)] {};
      \node [dashedbox, fit=(global_model)] {};
    
    \end{tikzpicture}
\caption{An example of vertical federated reinforcement learning (VFRL) architecture}
\end{figure}

\subsection{Advantages of FRL}
FRL offers several significant advantages over traditional RL approaches, particularly in distributed and privacy-sensitive applications. These advantages directly address critical challenges in implementing RL in real-world scenarios.

\subsubsection{Privacy Preservation with Effective Information Exchange}
One of the primary advantages of FRL is its ability to facilitate information exchange between agents while protecting privacy. The paper states that "FL can not only complete information exchange while avoiding privacy disclosure, but also adapt various agents to their different environments" \cite{qi2021federated}. This privacy-preserving mechanism is essential in applications where agent data may contain sensitive information that should not be shared directly with other participants or even with a central server.

\subsubsection{Bridging the Simulation-Reality Gap}
A significant challenge in traditional RL is the disparity between simulated training environments and real-world deployment conditions. FRL offers a solution to this problem as it "can aggregate information from both environments and thus bridge the gap between them" \cite{qi2021federated}. By enabling collaboration between agents operating in both simulated and real environments, FRL creates more robust and adaptable policies that perform well in actual deployment scenarios.

\subsubsection{Addressing Sample Efficiency and Large State-Action Spaces}
FRL provides solutions to core RL challenges related to sample efficiency and large state-action spaces. The paper notes that "in the case of large action space and state space, the performance of agents is vulnerable to collected samples since it is nearly impossible to explore all sampling spaces" \cite{qi2021federated}. By enabling agents to share learning experiences through model parameters rather than raw data, FRL significantly improves sample efficiency and accelerates the learning process.

\subsubsection{Integration of Partial Observations}
In many practical scenarios, individual agents have access to only partial observations of the environment. FRL addresses this limitation by enabling the integration of these partial observations through secure aggregation. The paper highlights that "in some cases, only partial features can be observed by each agent in RL... FL makes it possible to integrate this information through aggregation" \cite{qi2021federated}. This capability allows for more comprehensive decision-making based on collective information without requiring direct data sharing.


\subsection{Federated RL Communication Structures}

\subsubsection{Star Communication (Centralized Aggregation)}
Star communication follows a client-server architecture, where a central aggregator (server) collects model updates from multiple agents, processes them, and distributes the updated model back to all agents \cite{jin2022federated, liu2021federated}.

In this architecture, each agent independently trains a local model using its unique dataset. Instead of directly exchanging model parameters with other agents, they send updates to a central server. The server performs model aggregation using techniques such as FedAvg \cite{zhuo2019federated, zhang2023advances}, where a weighted averaging method is applied to combine updates from all agents. The newly aggregated model is then redistributed to all agents, ensuring that each participant benefits from the collective learning process. This iterative process continues until the global model converges.

One of the main advantages of this approach is its scalability. The central server efficiently manages communication between agents, reducing network congestion. Additionally, since agents do not communicate with each other directly, privacy is preserved because only model updates are shared rather than raw data \cite{liu2021federated}. This setup ensures that all agents follow a unified learning trajectory, leading to consistent policy updates and stability in learning \cite{zhang2023advances}.

However, this approach has its limitations. The central server becomes a single point of failure, meaning if it crashes or becomes compromised, the entire learning process halts \cite{liu2021federated}. Furthermore, as the number of agents increases, the server can experience bottlenecks, slowing down aggregation and increasing latency \cite{zhang2023advances}. Lastly, since all agents receive the same aggregated model, they may struggle to adapt to highly heterogeneous environments where local variations are significant.

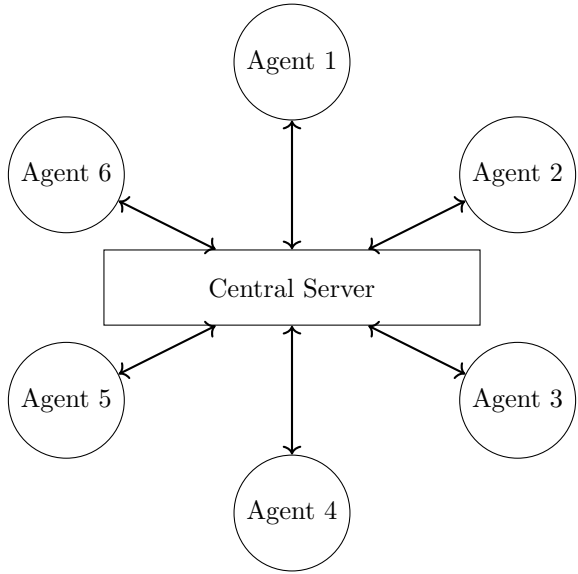
\begin{center}
\begin{tikzpicture}[scale=1, every node/.style={draw, circle, minimum size=1cm}, node distance=3.5cm]
    \node (S) at (0,0) [rectangle, draw, minimum width=5cm, minimum height=1cm] {Central Server};
    \node (A) at (0,3) {Agent 1};
    \node (B) at (3,1.5) {Agent 2};
    \node (C) at (3,-1.5) {Agent 3};
    \node (D) at (0,-3) {Agent 4};
    \node (E) at (-3,-1.5) {Agent 5};
    \node (F) at (-3,1.5) {Agent 6};
    \draw[<->, thick] (A) -- (S);
    \draw[<->, thick] (B) -- (S);
    \draw[<->, thick] (C) -- (S);
    \draw[<->, thick] (D) -- (S);
    \draw[<->, thick] (E) -- (S);
    \draw[<->, thick] (F) -- (S);
\end{tikzpicture}
\end{center}

\subsubsection{All-to-All Communication (Decentralized Aggregation)}
All-to-all communication eliminates the need for a central aggregator. Instead, agents exchange model updates directly with their peers in a distributed network topology \cite{zhuo2019federated, liu2021federated}.

Each agent trains a local model using its dataset and periodically communicates with selected peers, rather than relying on a central server. Model updates are shared among neighboring agents and aggregated using decentralized consensus-based techniques such as decentralized averaging \cite{jin2022federated, zhang2023advances}. The process continues iteratively, with each agent incorporating the updates received from its peers to refine its model until convergence is reached.

One of the primary benefits of all-to-all communication is its robustness. Since there is no central authority, the system can continue operating even if some agents drop out, making it highly fault-tolerant \cite{liu2021federated}. This approach is particularly beneficial in dynamic environments where agents operate under different conditions. Additionally, privacy is enhanced because model updates are exchanged only between direct peers rather than a central server \cite{zhuo2019federated}. This setup is well-suited for applications such as distributed sensor networks and multi-agent robotic systems.

However, this method also presents challenges. Communication overhead is significantly higher compared to star communication because each agent must exchange updates with multiple peers, leading to increased bandwidth consumption \cite{zhang2023advances}. Furthermore, maintaining synchronization between all agents is difficult due to differences in local processing speeds and network conditions \cite{liu2021federated}. Additionally, achieving convergence in decentralized networks can be slower compared to centralized aggregation, as updates propagate gradually across the network rather than being centrally applied at once.

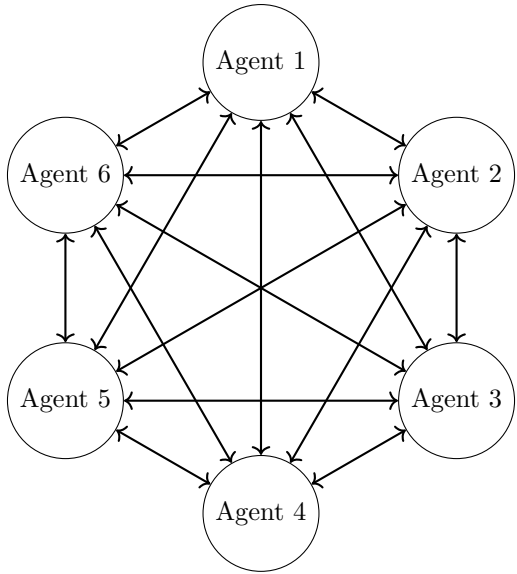
\begin{center}
\begin{tikzpicture}[scale=1, every node/.style={draw, circle, minimum size=1cm}, node distance=3.5cm]
    \node (A) at (0,3) {Agent 1};
    \node (B) at (2.6,1.5) {Agent 2};
    \node (C) at (2.6,-1.5) {Agent 3};
    \node (D) at (0,-3) {Agent 4};
    \node (E) at (-2.6,-1.5) {Agent 5};
    \node (F) at (-2.6,1.5) {Agent 6};
    \draw[<->, thick] (A) -- (B);
    \draw[<->, thick] (A) -- (C);
    \draw[<->, thick] (A) -- (D);
    \draw[<->, thick] (A) -- (E);
    \draw[<->, thick] (A) -- (F);
    \draw[<->, thick] (B) -- (C);
    \draw[<->, thick] (B) -- (D);
    \draw[<->, thick] (B) -- (E);
    \draw[<->, thick] (B) -- (F);
    \draw[<->, thick] (C) -- (D);
    \draw[<->, thick] (C) -- (E);
    \draw[<->, thick] (C) -- (F);
    \draw[<->, thick] (D) -- (E);
    \draw[<->, thick] (D) -- (F);
    \draw[<->, thick] (E) -- (F);
\end{tikzpicture}
\end{center}


\subsection{Federated Reinforcement Learning Algorithms}

The following algorithms are based on the Federated Reinforcement Learning (FedRL) framework introduced by Jin et al. in their paper "Federated Reinforcement Learning with Environment Heterogeneity" \cite{jin2022federated}. This framework addresses scenarios where multiple agents operate in environments with different state transition dynamics while sharing the same state space, action space, and reward function. The main challenge is collaborative learning without sharing raw trajectory data to preserve privacy. The objective is formalized as optimizing a policy that maximizes the average expected return across all environments:

\begin{equation}
g_{d_0}(\pi) = \frac{1}{n} \sum_{i=1}^{n} \mathbb{E} \left[ \sum_{t=1}^{\infty} \gamma^t R(s_t, a_t) | s_0 \sim d_0, a_t \sim \pi(\cdot|s_t), s_{t+1} \sim P_i(\cdot|s_t, a_t) \right]
\end{equation}

Where $d_0$ represents the initial state distribution, $\pi$ is the policy being optimized, and $P_i$ represents the transition dynamics of environment $i$. Jin et al. prove that no single policy can be simultaneously optimal across all heterogeneous environments, as the optimal policy depends on the initial state distribution.

\subsubsection{QAvg Algorithm: Federated Q-Learning}

QAvg is a federated version of Q-learning, where agents train local Q-values and periodically aggregate them at a server \cite{jin2022federated}. Each agent updates its local Q-function using its environment's dynamics. The update combines a standard Q-learning update with a momentum term $(1-\eta_t)$ that helps stabilize learning across heterogeneous environments. After several local updates, the server averages the Q-tables from all agents and redistributes this averaged Q-table back to each agent. Only the Q-values are communicated, preserving the privacy of each agent's experiences.

\begin{algorithm}[H]
\caption{QAvg Algorithm (Federated Q-Learning)\cite{jin2022federated}}
\begin{algorithmic}[1]
\State \textbf{Initialize:} Each agent $k$ initializes $Q^k_0(s,a)$ arbitrarily for all states $s$ and actions $a$
\For{each round $t = 0, 1, 2, \ldots$}
    \State \textbf{Local Updates:} Each agent $k$ performs multiple local updates:
    \State $Q^k_{t+1}(s, a) \leftarrow (1 - \eta_t) \cdot Q^k_t(s, a) + \eta_t \cdot \left[ R(s, a) + \gamma \sum_{s'} P_k(s' | s, a) \max_{a'} Q^k_t(s', a') \right]$
    \State \textbf{Global Aggregation:} The server performs:
    \State $\bar{Q}_t(s, a) \leftarrow \frac{1}{n} \sum_{k=1}^{n} Q^k_t(s, a)$, for all $s, a$
    \State \textbf{Distribution:} For all agents $k$ and all state-action pairs:
    \State $Q^k_t(s, a) \leftarrow \bar{Q}_t(s, a)$
\EndFor
\end{algorithmic}
\end{algorithm}

Jin et al. theoretically prove that QAvg converges to a suboptimal solution where the degree of suboptimality depends on the heterogeneity of the environments. The algorithm is most effective when the state transition functions across environments are similar.

\subsubsection{PAvg Algorithm: Federated Policy Gradient}

PAvg extends policy gradient methods to the federated setting, allowing agents to update local policies while aggregating globally \cite{jin2022federated}. Each agent independently performs policy gradient updates based on its local environment. After computing the gradient, the policy is projected back onto the action probability simplex to ensure it remains a valid probability distribution. Following several local updates, the server aggregates the policies from all agents and distributes the averaged policy back. Similar to QAvg, only the policy functions are shared, not the raw trajectories.

\begin{algorithm}[H]
\caption{PAvg Algorithm (Federated Policy Gradient)\cite{jin2022federated}}
\begin{algorithmic}[1]
\State \textbf{Initialize:} Each agent $k$ initializes policy $\pi^k_0(a|s)$ for all states $s$ and actions $a$
\For{each round $t = 0, 1, 2, \ldots$}
    \State \textbf{Local Updates:} Each agent $k$ performs multiple local updates:
    \State $\tilde{\pi}^k_{t+1}(a|s) \leftarrow \pi^k_t(a|s) + \frac{\partial g_{d_0,k}(\pi^k_t)}{\partial \pi(a|s)}$, for all $s, a$
    \State $\pi^k_{t+1}(\cdot|s) \leftarrow \text{Proj}_{\Delta(A)}(\tilde{\pi}^k_{t+1}(\cdot|s))$, for all $s$
    \State \textbf{Global Aggregation:} The server performs:
    \State $\bar{\pi}_t(a|s) \leftarrow \frac{1}{n} \sum_{k=1}^{n} \pi^k_t(a|s)$, for all $s, a$
    \State \textbf{Distribution:} For all agents $k$ and all state-action pairs:
    \State $\pi^k_t(a|s) \leftarrow \bar{\pi}_t(a|s)$
\EndFor
\end{algorithmic}
\end{algorithm}

The theoretical analysis by Jin et al. shows that PAvg also converges to a suboptimal solution with suboptimality dependent on environment heterogeneity. The algorithm provides a direct way to optimize policies when the optimal policy structure is known or when policy-based methods are preferred.

\subsubsection{DQNAvg: Federated Deep Q-Networks}

DQNAvg extends QAvg to deep reinforcement learning scenarios where the state space is too large for tabular representations \cite{jin2022federated}. Each agent maintains a local Q-network that is updated using standard DQN techniques including experience replay and target networks. Instead of sharing Q-tables as in QAvg, agents share the parameters of their neural networks. The server periodically aggregates these parameters and redistributes them to all agents.

\begin{algorithm}[H]
\caption{DQNAvg (Federated Deep Q-Networks)\cite{jin2022federated}}
\begin{algorithmic}[1]
\State \textbf{Initialize:} Each agent $k$ initializes Q-network parameters $\theta^k$
\For{each round $t = 0, 1, 2, \ldots$}
    \State \textbf{Local Updates:} Each agent $k$ performs multiple DQN updates:
    \State Collect experiences $(s, a, r, s')$ by interacting with environment $k$
    \State Update Q-network using standard DQN techniques:
    \State $\theta^k \leftarrow \theta^k - \alpha \nabla_{\theta^k} \left( r + \gamma \max_{a'} Q_{\theta^k}(s', a') - Q_{\theta^k}(s, a) \right)^2$
    \State \textbf{Global Aggregation:} The server averages network parameters:
    \State $\theta_{\text{global}} \leftarrow \frac{1}{n} \sum_{k=1}^{n} \theta^k$
    \State \textbf{Distribution:} For all agents $k$:
    \State $\theta^k \leftarrow \theta_{\text{global}}$
\EndFor
\end{algorithmic}
\end{algorithm}

This approach enables FedRL to scale to more complex environments with large or continuous state spaces that cannot be effectively represented in tabular form. The neural network architecture allows for function approximation and better generalization across states.

\subsubsection{DDPGAvg: Federated Deep Deterministic Policy Gradient}

DDPGAvg extends PAvg to continuous action spaces using the Deep Deterministic Policy Gradient algorithm \cite{jin2022federated}. Each agent maintains local actor and critic networks that are updated using standard DDPG techniques. After several local updates, the server aggregates the parameters of both networks across all agents and redistributes them.

\begin{algorithm}[H]
\caption{DDPGAvg (Federated Deep Deterministic Policy Gradient)\cite{jin2022federated}}
\begin{algorithmic}[1]
\State \textbf{Initialize:} Each agent $k$ initializes actor network $\theta^{\pi^k}$ and critic network $\theta^{Q^k}$
\For{each round $t = 0, 1, 2, \ldots$}
    \State \textbf{Local Updates:} Each agent $k$ performs multiple DDPG updates:
    \State Collect experiences $(s, a, r, s')$ by interacting with environment $k$
    \State Update critic network:
    \State $\theta^{Q^k} \leftarrow \theta^{Q^k} - \alpha_Q \nabla_{\theta^{Q^k}} \left( r + \gamma Q_{\theta^{Q^k}}(s', \pi_{\theta^{\pi^k}}(s')) - Q_{\theta^{Q^k}}(s, a) \right)^2$
    \State Update actor network:
    \State $\theta^{\pi^k} \leftarrow \theta^{\pi^k} + \alpha_{\pi} \nabla_{\theta^{\pi^k}} Q_{\theta^{Q^k}}(s, \pi_{\theta^{\pi^k}}(s))$
    \State \textbf{Global Aggregation:} The server averages network parameters:
    \State $\theta^{\pi}_{\text{global}} \leftarrow \frac{1}{n} \sum_{k=1}^{n} \theta^{\pi^k}$
    \State $\theta^{Q}_{\text{global}} \leftarrow \frac{1}{n} \sum_{k=1}^{n} \theta^{Q^k}$
    \State \textbf{Distribution:} For all agents $k$:
    \State $\theta^{\pi^k} \leftarrow \theta^{\pi}_{\text{global}}$
    \State $\theta^{Q^k} \leftarrow \theta^{Q}_{\text{global}}$
\EndFor
\end{algorithmic}
\end{algorithm}

This algorithm is particularly useful for continuous control tasks where the action space is continuous rather than discrete. The actor network directly outputs actions rather than action probabilities, and the critic network provides value estimates to guide the actor's updates.

\subsubsection{Personalized Federated RL (PerDQNAvg \& PerDDPGAvg)}

Recognizing the limitations of a one-size-fits-all approach, Jin et al. propose personalized versions of their algorithms through environment embeddings \cite{jin2022federated}. During training, agents share all network parameters except their embedding layers, allowing them to learn a common policy structure while retaining environment-specific information.

\begin{algorithm}[H]
\caption{Personalized Federated RL (PerDQNAvg \& PerDDPGAvg)\cite{jin2022federated}}
\begin{algorithmic}[1]
\State \textbf{Initialize:} Each agent $k$ initializes network parameters $\theta^k$ and environment embedding $e^k$
\For{each round $t = 0, 1, 2, \ldots$}
    \State \textbf{Local Updates:} Each agent $k$ performs multiple updates:
    \State Collect experiences $(s, a, r, s')$ by interacting with environment $k$
    \State Update network using standard techniques, conditioning on environment embedding $e^k$:
    \State $\pi^k(a | s, e^k) = \pi_{\theta}(a | s, e^k)$
    \State Update embedding vector using gradient descent:
    \State $e^k \leftarrow e^k - \eta \nabla_e J(e)$
    \State \textbf{Global Aggregation:} The server averages all parameters except embeddings:
    \State $\theta_{\text{global}} \leftarrow \frac{1}{n} \sum_{k=1}^{n} \theta^k$ (excluding embedding layers)
    \State \textbf{Distribution:} For all agents $k$:
    \State $\theta^k \leftarrow \theta_{\text{global}}$ (keeping original embedding $e^k$)
\EndFor
\end{algorithmic}
\end{algorithm}

This approach enables better performance in individual environments while still benefiting from collaborative learning. Additionally, it facilitates generalization to new environments by only requiring adjustment of the embedding layer rather than retraining the entire model.
\subsection{Theoretical Results}
From \cite{jin2022federated}, the results pertain to the behavior of the proposed algorithms QAvg and PAvg in federated reinforcement learning (FedRL) under heterogeneous environments.

\subsubsection*{Theorem 1: Dependence of Optimal Policy on Initial State Distribution}

\textbf{Theorem 1.}  
There exists a task of FedRL with the following properties. Assume that $\pi^\star \in \arg\max_\pi g_{d_0}(\pi)$. There exists another initial state distribution $d'_0$ and another policy $\tilde{\pi}$ such that:
\[
g_{d_0}(\tilde{\pi}) < g_{d_0}(\pi^\star), \quad \text{but} \quad g_{d'_0}(\tilde{\pi}) > g_{d'_0}(\pi^\star).
\]

\textit{Explanation:} This theorem demonstrates that in FedRL with heterogeneous environments, there does not exist a single optimal policy $\pi^\star$ that is universally optimal across all initial state distributions. The optimal policy depends on the specific initial state distribution $d_0$.

\subsubsection*{Theorem 2: Convergence to Suboptimal Solutions}

\textbf{Theorem 2.}  
Both QAvg and PAvg converge to suboptimal solutions, and the suboptimality is affected by the degree of environment heterogeneity in FedRL.

\textit{Explanation:} This result establishes that while QAvg and PAvg algorithms are guaranteed to converge, they converge to solutions that are suboptimal compared to environment-specific optimal policies. The extent of suboptimality is directly influenced by the heterogeneity of the environments.

\subsection{Summary}

\begin{flushleft}
\begin{tabular}{|p{1cm}|p{2cm}|p{3cm}|p{2.5cm}|p{2cm}|p{3cm}|p{2cm}|}
\hline
\textbf{Method} & \textbf{Type of RL Algorithm} & \textbf{Theoretical Guarantee} & \textbf{Key Properties} & \textbf{Comm Structure} & \textbf{Applicability} \\

\hline
QAvg & Value-based (Tabular Q-Learning) & Converges to a suboptimal solution; suboptimality depends on environment heterogeneity. & Aggregates Q-tables; preserves privacy by sharing only Q-values. & Star (Centralized Aggregation) & Discrete state-action spaces. \\
\hline

PAvg & Policy-based (Policy Gradient) & Converges to a suboptimal solution; suboptimality depends on environment heterogeneity. & Aggregates policy parameters; preserves privacy by sharing only policies. & Star (Centralized Aggregation) & Discrete or continuous action spaces. \\
\hline

DQNAvg & Value-based (Deep Q-Networks) & Extends QAvg to large/continuous state spaces; suboptimality due to heterogeneity. & Aggregates neural network parameters; uses experience replay and target networks. & Star (Centralized Aggregation) & Large or continuous state spaces. \\
\hline

DDPGAvg & Policy-based (DDPG) & Extends PAvg to continuous action spaces; suboptimality due to heterogeneity. & Aggregates actor-critic network parameters; suitable for continuous control tasks. & Star (Centralized Aggregation) & Continuous action spaces. \\
\hline
\end{tabular}

\begin{tabular}{|p{1cm}|p{2cm}|p{3cm}|p{2.5cm}|p{2cm}|p{3cm}|p{2cm}|}

\hline
\textbf{Method} & \textbf{Type of RL Algorithm} & \textbf{Theoretical Guarantee} & \textbf{Key Properties} & \textbf{Comm Structure} & \textbf{Applicability} \\

\hline

PDQNAvg & Personalized (DQN) & No explicit guarantees; improves performance in heterogeneous environments. & Shares all network parameters except embedding layers; retains environment-specific adaptations. & Star (Centralized Aggregation) & Environments requiring personalization. \\
\hline

PDDPGAvg & Personalized (DDPG) & No explicit guarantees; improves performance in heterogeneous environments. & Similar to PerDQNAvg but for continuous actions; combines shared policy structure with local embeddings. & Star (Centralized Aggregation) & Continuous actions with personalization. \\

HFRL & General Framework & Depends on underlying algorithm (e.g., QAvg, PAvg). & Agents share state-action spaces but have different experiences; aggregates Q-values or policies. & Star or All-to-All & Homogeneous state-action spaces. \\
\hline

VFRL & General Framework & Depends on underlying algorithm (e.g., DQNAvg, DDPGAvg). & Agents observe different features; integrates partial observations via feature or value aggregation. & Star or All-to-All & Heterogeneous or partitioned environments. \\
\hline

\end{tabular}
\end{flushleft}

\subsection{Open Problems and Future Work}

Federated Reinforcement Learning (FedRL) with environment heterogeneity represents a burgeoning field that combines the challenges of distributed learning with the complexities of diverse environmental dynamics. While recent advancements, such as the QAvg and PAvg algorithms proposed by Jin et al. \cite{jin2022federated}, have laid foundational frameworks for collaborative policy optimization across heterogeneous environments, significant open problems remain. This section synthesizes unresolved challenges and outlines promising directions for future research, drawing insights from theoretical analyses, algorithmic limitations, and practical considerations highlighted in the literature.

\subsection{Theoretical Foundations and Limitations}

The convergence guarantees of QAvg and PAvg algorithms reveal a fundamental dependency on environment heterogeneity in federated reinforcement learning systems. Jin et al. demonstrate that both approaches inevitably converge to suboptimal policies, with performance gaps directly proportional to the divergence between state transition dynamics across individual environments~\cite{jin2022federated}. This finding quantifies the inherent trade-off between collaborative learning and environmental heterogeneity, raising several critical questions about potential improvements to the current paradigm. Current aggregation mechanisms employ naive averaging of local Q-tables or policies, which potentially dilutes environment-specific nuances that could be critical for optimal performance. Future research might explore more sophisticated weighted averaging schemes where aggregation weights dynamically adjust based on measured pairwise environment similarities or task affinities, potentially mitigating the heterogeneity-induced suboptimality that plagues current approaches.

The influence of initial state distributions further complicates the FedRL landscape. As demonstrated in Theorem 1 of Jin et al., the optimal policy in FedRL exhibits dependency on the initial state distribution $d_0$~\cite{jin2022federated}. This dependence significantly complicates policy generalization across environments with divergent starting conditions, necessitating more robust analytical frameworks that can account for these distribution shifts. The theoretical framework introduces an important construct-the imaginary environment $M_I$ with averaged transition dynamics $\bar{P}$-which serves as an analytical proxy for convergence analysis. However, this construct introduces its own limitations, as $M_I$ may not correspond to any actual real-world environment, thereby constraining the interpretability and practical applicability of the theoretical guarantees. Developing more formalized relationships between $M_I$ and the original environments could potentially yield tighter performance bounds and more actionable insights for practitioners implementing these systems.

The existing convergence analyses predominantly assume tabular representations of Q-functions and policies. This assumption diverges significantly from modern reinforcement learning applications, which overwhelmingly employ function approximators such as deep neural networks. While Jin et al. propose extended approaches like DQNAvg and DDPGAvg for non-tabular settings, the theoretical properties of these extensions remain largely uncharacterized~\cite{jin2022federated}. This gap raises fundamental questions about convergence conditions in heterogeneous settings when using neural network approximations. The relationship between network overparameterization and the critical trade-off between generalization and personalization remains unexplored, as does the potential for approximation errors introduced by neural architectures to amplify the suboptimality issues already present due to environment heterogeneity.

\subsection{Personalization Approaches and Challenges}

The personalization heuristic proposed by Jin et al.-embedding environments into low-dimensional vectors-demonstrates empirical performance improvements but currently lacks comprehensive theoretical justification~\cite{jin2022federated}. Bridging this theoretical gap requires formal characterization of the embedding spaces to understand what geometric or topological properties these environment embeddings should satisfy to effectively capture transition dynamics. These embeddings might be productively related to existing concepts in reinforcement learning literature, such as bisimulation metrics, providing a firmer foundation for the approach. Additionally, while the global policy converges to a suboptimal solution, the complex interaction between personalized embeddings and global network parameters remains unanalyzed. A two-timescale analysis framework, where embeddings evolve faster than global weights, could provide valuable insights into this relationship and guide more effective algorithm design.

The heuristic allows for fine-tuning embeddings when encountering new environments, but theoretical limits on this adaptability remain undefined. Developing information-theoretic bounds on the required interaction steps for adaptation could guide practical deployments and set appropriate expectations for performance in novel settings. Current personalization methods also implicitly assume alignment between local and global objectives, which may not hold in real-world applications where environment-specific rewards diverge significantly. Future algorithmic developments could incorporate negotiation mechanisms inspired by cooperative game theory to balance individual and collective goals in these multi-objective optimization scenarios.

Meta-learning frameworks represent another promising direction, where the federated model learns initialization parameters specifically designed to facilitate rapid personalization across diverse environments. Such approaches could be complemented by dynamic resource allocation strategies that prioritize training on environments where personalization yields the most significant marginal benefits to the global model. These advances would address the fundamental tension between personalization and generalization that characterizes federated reinforcement learning systems.

\subsection{Scalability, Communication, and Implementation Challenges}

The experiments documented in Jin et al. primarily focus on tabular and low-dimensional control tasks, leaving open questions about scaling FedRL to high-dimensional spaces such as pixel-based observations~\cite{jin2022federated}. This scaling introduces significant challenges related to the curse of dimensionality, as aggregating high-dimensional policies or value functions may require sophisticated compression techniques like knowledge distillation or network pruning to remain computationally feasible. The interaction between environment heterogeneity and partial observability represents another unexplored frontier, as federated variants of POMDP algorithms have received minimal attention in the literature thus far. While DDPGAvg has been proposed for continuous action spaces, its efficacy in heterogeneous continuous control tasks such as multi-robot systems with varying dynamics requires more rigorous evaluation before deployment in complex domains.

Frequent model aggregation in FedRL introduces substantial communication costs, particularly when utilizing large neural networks. This challenge has prompted exploration of several potential solutions to improve communication efficiency. Adaptive communication schedules could allow agents to update their models locally until a performance plateau is detected, thereby reducing redundant transmissions without compromising learning progress. Delta parameter synchronization, which transmits only model parameter changes rather than full weights, has shown promise in federated supervised learning and could be adapted to the reinforcement learning context. Federated ensemble methods offer yet another approach, training smaller specialized models for each environment and aggregating them through ensemble techniques such as Bayesian committee machines.

\subsection{Privacy, Security, and Ethical Considerations}

While Jin et al. prevent explicit trajectory sharing to preserve privacy, sophisticated adversaries might still infer environment properties from the aggregated models~\cite{jin2022federated}. Strengthening privacy guarantees requires more robust mechanisms. Differential privacy (DP) approaches involving noise injection during aggregation or local updates could enhance protection, though this may exacerbate the suboptimality issues already present in FedRL systems. Quantifying the precise trade-off between differential privacy guarantees and utility in the FedRL context represents a critical area for future research. Secure multi-party computation (SMPC) offers an alternative approach, implementing cryptographic protocols to perform aggregation without revealing individual model updates, though these come with significant computational overhead that must be carefully managed.

The distinction between cross-device and cross-silo federated RL introduces additional complexities. Most current FedRL research assumes cross-device settings with numerous simple agents, but cross-silo scenarios-involving fewer but more capable agents, such as autonomous vehicles or industrial robots-present distinct challenges. Heterogeneous computing capabilities across these agents necessitate asynchronous aggregation protocols to synchronize training effectively. Non-IID environment distributions in cross-silo settings, which may exhibit systematic heterogeneity due to factors like regional differences, could benefit from clustered aggregation strategies that recognize and leverage these patterns.

Deploying FedRL policies from simulation to physical systems introduces the challenge of accounting for real-world heterogeneity not fully captured during training. Promising approaches to address this sim-to-real gap include domain randomization within federated training, explicitly sampling environment parameters from broader distributions to improve robustness. Additionally, online adaptation protocols could continuously incorporate real-world interaction data while preserving privacy constraints, allowing for ongoing refinement of models in deployment settings. In open federated systems with self-interested agents, participation may be limited if collaboration offers minimal local benefits. Economic frameworks could address this through mechanisms that price personalized model improvements, allowing agents to "purchase" better personalization through increased participation in the federated learning process.

The environmental impact of training large-scale FedRL systems is another important consideration, as these processes require significant energy consumption. Research into energy-efficient federated training through sparser communications, model sparsification, or specialized green computing techniques represents an imperative direction for responsible advancement of the field.

\subsection{Synthesis and Future Directions}

Federated reinforcement learning represents a powerful paradigm for collaborative learning across distributed agents with private experiences, but current approaches face significant limitations in theoretical foundations, personalization mechanisms, scalability, and privacy guarantees. Advancing beyond naive averaging techniques toward more sophisticated aggregation mechanisms that account for environment similarities could significantly reduce heterogeneity-induced suboptimality. Similarly, developing stronger theoretical understandings of embedding-based personalization approaches would enable more principled algorithm design and performance guarantees across diverse environments.

As the field moves toward high-dimensional observation spaces and more complex tasks, addressing communication efficiency and scalability becomes increasingly critical. Privacy and security considerations must evolve in parallel with algorithmic advances to ensure that the benefits of federated learning-particularly the privacy preservation of local trajectories-are not undermined by sophisticated inference attacks on aggregated models. Finally, ethical considerations around incentive design, and environmental impact should inform research directions to ensure that federated reinforcement learning technologiesmitigate benefit diverse stakeholders while minimizing potential harms.

By addressing these challenges through a coordinated research agenda, the federated reinforcement learning community can develop more robust, efficient, and equitable approaches to collaborative learning across heterogeneous environments, ultimately unlocking the full potential of this promising paradigm for applications ranging from robotics and autonomous vehicles to personalized healthcare and smart infrastructure systems.

\section{Cooperative Decentralized RL}

\subsection{Introduction}

Cooperative Decentralized Reinforcement Learning (CDRL) is an emerging research area that combines the principles of decentralized learning with the strategic framework of multi-agent reinforcement learning (MARL). In CDRL, multiple agents learn to act cooperatively without relying on a centralized controller, instead making decisions based on local observations and limited communication with nearby peers. This paradigm is particularly relevant for real-world applications where centralized control is impractical due to communication constraints, scalability issues, or privacy concerns. In decentralized learning, each agent maintains its own local policy and value estimates, yet must collaborate to achieve a shared objective. The inherent challenge lies in coordinating learning across distributed nodes with partial information, which often requires designing robust communication protocols and consensus mechanisms. MARL provides the formal framework for this setting by modeling the interactions among multiple agents, whose joint actions influence the environment and the resulting rewards. Cooperative MARL, a subset of this framework, specifically focuses on scenarios where all agents share a common goal, such as maximizing a global return. In these settings, the agents are incentivized to coordinate their actions, yet they must do so in a manner that preserves decentralized operation. The work by \cite{zhang2018n} represents a significant advancement in CDRL by addressing these challenges with a fully decentralized approach that leverages consensus updates and decentralized actor–critic algorithms.

In the following sections, we review the contributions of \cite{zhang2018n}'s paper in detail, examining how their methods effectively address the challenges of cooperative decentralized reinforcement learning.

\subsection{Zhang et al (2018) Review}
\label{sec:intro}
\cite{zhang2018n} propose a framework that enables agents to operate autonomously without the need for a centralized controller. In this setup, each agent is designed to independently process a local reward—which may differ from agent to agent—while determining its actions through an individualized, parameterized policy. Rather than relying on a global communication scheme, agents share information only with their immediate neighbors over a network whose connectivity may vary over time. Despite this variability, the system is structured such that the communication network remains jointly connected over bounded time intervals, ensuring that every agent can indirectly interact with all others.

The global objective is to maximize the \emph{globally averaged return} defined as
\[
J(\theta) = \sum_{s \in \mathcal{S}} d_{\theta}(s) \sum_{a \in \mathcal{A}} \pi_{\theta}(s,a) R(s,a),
\]
where \(d_{\theta}(s)\) is the stationary state distribution under the joint policy \(\pi_{\theta}\) and 
\[
R(s,a) = \frac{1}{N}\sum_{i=1}^N R_i(s,a)
\]
is the averaged reward over \(N\) agents. Importantly, while the communication network is time-varying and need not be connected at every time step, it is assumed to be \emph{jointly connected over time}---i.e., there exists a bounded time interval \(T\) within which every agent can indirectly communicate with every other agent.

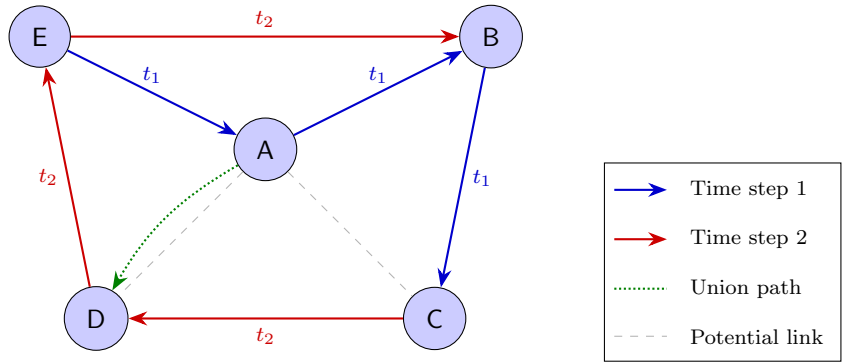
\begin{figure}[htbp]
    \centering
    \begin{tikzpicture}[
        scale=1.5,
        node distance=2.5cm, 
        agent/.style={circle, draw, fill=blue!20, inner sep=5pt, font=\sffamily},
        time1/.style={-{Stealth[length=2.5mm]}, thick, draw=blue!80!black},
        time2/.style={-{Stealth[length=2.5mm]}, thick, draw=red!80!black},
        inactive/.style={dashed, draw=gray!60},
        union/.style={densely dotted, thick, draw=green!50!black}
    ]
       
        \node[agent] (A) at (0,0) {A};
        \node[agent] (B) at (2,1) {B};
        \node[agent] (C) at (1.5,-1.5) {C};
        \node[agent] (D) at (-1.5,-1.5) {D};
        \node[agent] (E) at (-2,1) {E};

        \draw[time1] (A) -- (B) node[midway, above, font=\footnotesize, text=blue!80!black] {$t_1$};
        \draw[time1] (B) -- (C) node[midway, right, font=\footnotesize, text=blue!80!black] {$t_1$};
        \draw[time1] (E) -- (A) node[midway, above, font=\footnotesize, text=blue!80!black] {$t_1$};

        \draw[time2] (C) -- (D) node[midway, below, font=\footnotesize, text=red!80!black] {$t_2$};
        \draw[time2] (D) -- (E) node[midway, left, font=\footnotesize, text=red!80!black] {$t_2$};
        \draw[time2] (E) -- (B) node[midway, above, font=\footnotesize, text=red!80!black] {$t_2$};

        \draw[inactive] (A) -- (C);
        \draw[inactive] (A) -- (D);

        \draw[union, -{Stealth[length=2.5mm]}] (A) to[bend right=15] (D);
        
        \matrix[draw, column sep=5pt, row sep=5pt, nodes={anchor=west}, right] at (3,-1) {
            \draw[time1] (0,0) -- (0.7,0); & \node[font=\footnotesize]{Time step 1}; \\
            \draw[time2] (0,0) -- (0.7,0); & \node[font=\footnotesize]{Time step 2}; \\
            \draw[union] (0,0) -- (0.7,0); & \node[font=\footnotesize]{Union path}; \\
            \draw[inactive] (0,0) -- (0.7,0); & \node[font=\footnotesize]{Potential link}; \\
        };
    \end{tikzpicture}
    \caption{A dynamic communication topology illustrating time-varying connectivity among agents. 
    Direct links are active at different time steps ($t_1$ and $t_2$), while dashed lines represent potential 
    connections. The dotted green path shows an example of indirect connectivity achieved through the union of 
    connections over time.}
    \label{fig:graph_topology}
\end{figure} 

\subsubsection{Single-Agent MDP and Actor--Critic Methods}

A standard Markov decision process (MDP) is defined by the tuple \((\mathcal{S},\mathcal{A},P,r)\), where \(\mathcal{S}\) represents a finite state space, \(\mathcal{A}\) a finite action space, \(P(s'|s,a)\) the state transition probability, and \(r(s,a)\) the expected reward for taking action \(a\) in state \(s\). The objective in such a setting is to maximize the long-term average reward given by

\[
J(\pi) = \sum_{s\in\mathcal{S}} d_{\pi}(s) \sum_{a\in\mathcal{A}} \pi(s,a) r(s,a),
\]

where \(d_{\pi}(s)\) denotes the stationary distribution under the policy \(\pi\).

Actor--critic (AC) methods decompose the problem into:
\begin{enumerate}[label=(\alph*)]
    \item A \emph{critic} that estimates the value (or action-value) function via function approximation.
    \item An \emph{actor} that updates the policy parameters through gradient ascent.
\end{enumerate}
The classical policy gradient theorem states that
\[
\nabla_{\theta} J(\theta) = \mathbb{E}_{s\sim d_{\theta},\, a\sim\pi_{\theta}} \left[\nabla_{\theta}\log\pi_{\theta}(s,a)\, A_{\theta}(s,a)\right],
\]
with the advantage function \(A_{\theta}(s,a)= Q_{\theta}(s,a)-V_{\theta}(s)\).

\subsubsection{Networked Multi-Agent MDP (NM-MDP)}

\cite{zhang2018n} extend the single-agent Markov Decision Process (MDP) to a multi-agent framework involving \(N\) agents, which they refer to as the Networked Multi-Agent MDP (NM-MDP). In this setting, each agent \(i\) operates within its own local action space \(A_i\) and adheres to a local policy \(\pi_{\theta^i}^i(s,a^i)\). The overall joint policy is constructed by the factorization
\[
\pi_{\theta}(s,a) = \prod_{i=1}^{N} \pi_{\theta^i}^i(s,a^i),
\]
with the joint parameter vector defined as \(\theta = [\theta^1,\dots,\theta^N]\). Under this joint policy, the system's behavior is characterized by the stationary state distribution \(d_{\theta}(s)\) and an averaged reward given by
\[
R(s,a) = \frac{1}{N}\sum_{i=1}^{N} R_i(s,a).
\]

Key assumptions in the NM-MDP framework include several conditions to ensure theoretical soundness. The communication graph \(\mathcal{G}_t=(\mathcal{N},\mathcal{E}_t)\) is allowed to vary over time, yet it is assumed to be jointly connected over time so that every agent can eventually interact with any other despite intermittent disconnections. Moreover, the induced Markov chain under any joint policy \(\pi_{\theta}\) is assumed to be ergodic, ensuring long-term statistical regularity. The instantaneous rewards \(r_t^i\) are uniformly bounded to avoid issues with unbounded variations. The consensus weight matrices \(C_t = [c_t(i,j)]\) satisfy the condition
\[
\sum_{j\in\mathcal{N}} c_t(i,j)=1 \quad \forall\, i,
\]
with \(c_t(i,j)=0\) whenever \((i,j)\notin\mathcal{E}_t\). In addition, in expectation, these matrices are column-stochastic, and the spectral norm of
\[
E\left[C_t^\top \left(I-\frac{1}{N}\mathbf{11}^\top\right) C_t\right]
\]
is strictly less than one.

\subsubsection{Decentralized Policy Gradient Theorem}
\label{sec:pg_theorem}

\cite{zhang2018n} derive a decentralized policy gradient theorem using the log-derivative trick. Starting from the standard policy gradient,
\[
\nabla_{\theta} J(\theta) = \mathbb{E}_{s,a}\left[\nabla_{\theta}\log\pi_{\theta}(s,a) \, Q_{\theta}(s,a)\right],
\]
and noting the factorization
\[
\pi_{\theta}(s,a)=\prod_{i=1}^N \pi_{\theta^i}^i(s,a^i),
\]
one obtains for each agent \(i\):
\[
\nabla_{\theta^i}J(\theta) = \mathbb{E}_{s,a}\left[\nabla_{\theta^i}\log\pi_{\theta^i}^i(s,a^i) \, Q_{\theta}(s,a)\right].
\]
By introducing the \emph{local advantage function}
\[
A_{\theta}^i(s,a) = Q_{\theta}(s,a) - \widetilde{V}_{\theta}^i(s,a^{-i}),
\]
with
\[
\widetilde{V}_{\theta}^i(s,a^{-i}) = \sum_{a^i \in A_i} \pi_{\theta^i}^i(s,a^i)\, Q_{\theta}(s,a^i,a^{-i}),
\]
the gradient becomes
\[
\nabla_{\theta^i}J(\theta) = \mathbb{E}_{s\sim d_{\theta},\,a\sim\pi_{\theta}}\left[\nabla_{\theta^i}\log\pi_{\theta^i}^i(s,a^i) \, A_{\theta}^i(s,a)\right].
\]
In practice, each agent obtains a local estimate of \(Q_{\theta}(s,a)\) via consensus updates in the critic, ensuring that all agents eventually agree on a common approximation.

\subsubsection{Decentralized Actor--Critic Algorithms}
\label{sec:algorithms}

\cite{zhang2018n} propose two actor--critic algorithms that use consensus-based critic updates and two-time-scale updates.

\subsubsection*{Algorithm 1: Actor--Critic with Action-Value Function Approximation}
Each agent approximates the global action-value function using linear function approximation:
\[
Q(s,a;\omega)=\omega^\top \phi(s,a),
\]
where \(\phi(s,a)\) is the feature vector. The critic parameters are updated locally and then fused via consensus with weight matrix \(C_t\).

\paragraph{Key Steps:}
\begin{itemize}
    \item \textbf{Local Reward and Running Average:} Each agent \(i\) maintains an estimate \(\mu_i\) of its average reward.
    \item \textbf{Critic Update:} For agent \(i\), the temporal-difference error is computed as
    \[
    \delta_i^t = r_{t+1}^i - \mu_i^t + Q(s_{t+1},a_{t+1};\omega_i^t) - Q(s_t,a_t;\omega_i^t).
    \]
    A local critic update is performed:
    \[
    \widetilde{\omega}_i^t = \omega_i^t + \beta_{\omega,t}\,\delta_i^t\,\nabla_{\omega} Q(s_t,a_t;\omega_i^t).
    \]
    \item \textbf{Consensus Update:} Each agent updates its critic parameters via
    \[
    \omega_i^{t+1} = \sum_{j\in\mathcal{N}} c_t(i,j)\,\widetilde{\omega}_j^t.
    \]
    \item \textbf{Actor Update:} The local advantage is calculated as
    \[
    A_i^t = Q(s_t,a_t;\omega_i^t) - \sum_{a_i\in A_i} \pi_{\theta_i}^i(s_t,a_i)\,Q(s_t,a_i,a_{-i};\omega_i^t),
    \]
    and the policy update is performed using the score function
    \[
    \psi_i^t = \nabla_{\theta_i}\log \pi_{\theta_i}^i(s_t,a_i^t).
    \]
    The policy is then updated via
    \[
    \theta_i^{t+1} = \Gamma_i\Bigl(\theta_i^t + \beta_{\theta,t}\,A_i^t\,\psi_i^t\Bigr),
    \]
    where \(\Gamma_i\) is the projection operator onto the compact set \(\Theta_i\) (as per Assumption 4.1).
    \item \textbf{Policy Parameterization:} Although the paper permits general differentiable parameterizations, a common example is thesoftmax function:
    \[
    \pi_{\theta_i}^i(s,a_i) = \frac{\exp(f_{\theta_i}(s,a_i))}{\sum_{a'_i} \exp(f_{\theta_i}(s,a'_i))},
    \]
    where \(f_{\theta_i}(s,a_i)\) is differentiable.
\end{itemize}

\paragraph{Pseudocode:}

\begin{algorithm}[H]
\caption{Networked Actor--Critic with Action-Value Function Approximation}
\label{alg:alg1_final}
\begin{algorithmic}[1]
\State \textbf{Input:} Initial values \(\mu_i^0\), \(\omega_i^0\), \(\widetilde{\omega}_i^0\), \(\theta_i^0\) for all \(i\in\mathcal{N}\); initial state \(s_0\); stepsizes \(\{\beta_{\omega,t}\}\) and \(\{\beta_{\theta,t}\}\).
\State Each agent \(i\) executes \(a_i^0 \sim \pi_{\theta_i^0}^i(s_0,\cdot)\) and observes joint actions \(a_0=(a_1^0,\ldots,a_N^0)\).
\State Initialize \(t \leftarrow 0\).
\Repeat
    \For{each agent \(i \in \mathcal{N}\)}
        \State Observe \(s_{t+1}\) and reward \(r_{t+1}^i\).
        \State Update running average:
        \[
        \mu_i^{t+1} \leftarrow (1-\beta_{\omega,t})\mu_i^t + \beta_{\omega,t}\,r_{t+1}^i.
        \]
        \State Select action: \(a_i^{t+1} \sim \pi_{\theta_i^t}^i(s_{t+1},\cdot)\).
    \EndFor
    \State Observe joint actions \(a_{t+1}=(a_1^{t+1},\ldots,a_N^{t+1})\).
    \For{each agent \(i \in \mathcal{N}\)}
        \State Compute TD-error:
        \[
        \delta_i^t \leftarrow r_{t+1}^i - \mu_i^t + Q(s_{t+1},a_{t+1};\omega_i^t) - Q(s_t,a_t;\omega_i^t).
        \]
        \State \textbf{Critic Update:} 
        \[
        \widetilde{\omega}_i^t \leftarrow \omega_i^t + \beta_{\omega,t}\,\delta_i^t\,\nabla_{\omega} Q(s_t,a_t;\omega_i^t).
        \]
        \State Compute local advantage:
        \[
        A_i^t \leftarrow Q(s_t,a_t;\omega_i^t) - \sum_{a_i\in A_i} \pi_{\theta_i}^i(s_t,a_i)\,Q(s_t,a_i,a_{-i};\omega_i^t).
        \]
        \State Compute score function:
        \[
        \psi_i^t \leftarrow \nabla_{\theta_i} \log \pi_{\theta_i}^i(s_t,a_i^t).
        \]
        \State \textbf{Actor Update:} 
        \[
        \theta_i^{t+1} \leftarrow \Gamma_i\Bigl(\theta_i^t + \beta_{\theta,t}\,A_i^t\,\psi_i^t\Bigr).
        \]
        \State Send \(\widetilde{\omega}_i^t\) to neighbors \(\{j\in\mathcal{N}:(i,j)\in \mathcal{E}_t\}\).
    \EndFor
    \For{each agent \(i \in \mathcal{N}\)}
        \State \textbf{Consensus Step:}
        \[
        \omega_i^{t+1} \leftarrow \sum_{j\in\mathcal{N}} c_t(i,j)\,\widetilde{\omega}_j^t.
        \]
    \EndFor
    \State \(t \leftarrow t+1\).
\Until{Convergence}
\end{algorithmic}
\end{algorithm}

\subsubsection*{Algorithm 2: Actor--Critic with State-Value TD-Error}
In Algorithm 2, each agent employs a state-value function \(V(s;v)\) and a global reward estimator \(R(s,a;\lambda)\). Consensus updates are applied to the running average \(\mu_i\), the reward estimator \(\lambda_i\), and the critic parameter \(v_i\).

\paragraph{Key Steps:}
\begin{itemize}
    \item \textbf{Running Average and Reward Estimation:}
    \[
    \mu_{e,i}^t \leftarrow (1-\beta_{v,t})\mu_i^t + \beta_{v,t}\,r_{t+1}^i,
    \]
    \[
    \lambda_{e,i}^t \leftarrow \lambda_i^t + \beta_{v,t}\Bigl[r_{t+1}^i - R_t(\lambda_i^t)\Bigr]\nabla_{\lambda} R_t(\lambda_i^t).
    \]
    \item \textbf{Critic Update:} Compute the state-value TD-error:
    \[
    \delta_i^t \leftarrow r_{t+1}^i - \mu_i^t + V(s_{t+1};v_i^t) - V(s_t;v_i^t),
    \]
    and update:
    \[
    v_{e,i}^t \leftarrow v_i^t + \beta_{v,t}\,\delta_i^t\,\nabla_{v} V(s_t;v_i^t).
    \]
    \item \textbf{Actor Update:} Define the adjusted TD-error:
    \[
    \delta_{e,i}^t \leftarrow R_t(\lambda_i^t) - \mu_i^t + V(s_{t+1};v_i^t) - V(s_t;v_i^t),
    \]
    then update:
    \[
    \theta_i^{t+1} \leftarrow \Gamma_i\Bigl(\theta_i^t + \beta_{\theta,t}\,\delta_{e,i}^t\,\nabla_{\theta_i} \log \pi_{\theta_i}^i(s_t,a_i^t)\Bigr).
    \]
    \item \textbf{Consensus Updates:} Fuse the estimates via:
    \[
    \mu_i^{t+1} \leftarrow \sum_{j\in\mathcal{N}} c_t(i,j)\,\mu_{e,j}^t,\quad
    \lambda_i^{t+1} \leftarrow \sum_{j\in\mathcal{N}} c_t(i,j)\,\lambda_{e,j}^t,\quad
    v_i^{t+1} \leftarrow \sum_{j\in\mathcal{N}} c_t(i,j)\,v_{e,j}^t.
    \]
\end{itemize}

\paragraph{Pseudocode:}

\begin{algorithm}[H]
\caption{Networked Actor--Critic with State-Value TD-Error}
\label{alg:alg2_final}
\begin{algorithmic}[1]
\State \textbf{Input:} Initial values \(\mu_i^0\), \(\mu_{e,i}^0\), \(v_i^0\), \(v_{e,i}^0\), \(\lambda_i^0\), \(\lambda_{e,i}^0\), \(\theta_i^0\) for all \(i\in\mathcal{N}\); initial state \(s_0\); stepsizes \(\{\beta_{v,t}\}\) and \(\{\beta_{\theta,t}\}\).
\State Each agent \(i\) executes \(a_i^0 \sim \pi_{\theta_i^0}^i(s_0,\cdot)\).
\State Initialize \(t \leftarrow 0\).
\Repeat
    \For{each agent \(i \in \mathcal{N}\)}
        \State Observe \(s_{t+1}\) and reward \(r_{t+1}^i\).
        \State Update running average:
        \[
        \mu_{e,i}^t \leftarrow (1-\beta_{v,t})\mu_i^t + \beta_{v,t}\,r_{t+1}^i.
        \]
        \State Update reward estimator:
        \[
        \lambda_{e,i}^t \leftarrow \lambda_i^t + \beta_{v,t}\Bigl[r_{t+1}^i - R_t(\lambda_i^t)\Bigr]\nabla_{\lambda} R_t(\lambda_i^t).
        \]
        \State Compute state-value TD-error:
        \[
        \delta_i^t \leftarrow r_{t+1}^i - \mu_i^t + V(s_{t+1};v_i^t) - V(s_t;v_i^t).
        \]
        \State \textbf{Critic Update:}
        \[
        v_{e,i}^t \leftarrow v_i^t + \beta_{v,t}\,\delta_i^t\,\nabla_{v} V(s_t;v_i^t).
        \]
        \State Compute adjusted TD-error:
        \[
        \delta_{e,i}^t \leftarrow R_t(\lambda_i^t) - \mu_i^t + V(s_{t+1};v_i^t) - V(s_t;v_i^t).
        \]
        \State Compute score function:
        \[
        \psi_i^t \leftarrow \nabla_{\theta_i}\log\pi_{\theta_i}^i(s_t,a_i^t).
        \]
        \State \textbf{Actor Update:}
        \[
        \theta_i^{t+1} \leftarrow \Gamma_i\Bigl(\theta_i^t + \beta_{\theta,t}\,\delta_{e,i}^t\,\psi_i^t\Bigr).
        \]
        \State Send \(\mu_{e,i}^t\), \(\lambda_{e,i}^t\), \(v_{e,i}^t\) to neighbors over \(\mathcal{G}_t\).
    \EndFor
    \For{each agent \(i \in \mathcal{N}\)}
        \State \textbf{Consensus Step:}
        \[
        \mu_i^{t+1} \leftarrow \sum_{j\in\mathcal{N}} c_t(i,j)\,\mu_{e,j}^t,\quad
        \lambda_i^{t+1} \leftarrow \sum_{j\in\mathcal{N}} c_t(i,j)\,\lambda_{e,j}^t,\quad
        v_i^{t+1} \leftarrow \sum_{j\in\mathcal{N}} c_t(i,j)\,v_{e,j}^t.
        \]
    \EndFor
    \State \(t \leftarrow t+1\).
\Until{Convergence}
\end{algorithmic}
\end{algorithm}

\subsubsection{Convergence Analysis}
\label{sec:convergence_analysis}
 \cite{zhang2018n} establish the convergence of their decentralized learning algorithm under a series of standard assumptions. First, each agent's parameter vector, \(\theta_i\), is projected onto a compact set \(\Theta_i\) via the operator \(\Gamma_i\). This projection guarantees that the parameters remain bounded, thereby ensuring system stability. Additionally, the authors assume that the instantaneous rewards, \(r_t^i\), are uniformly bounded, which prevents divergence due to unbounded reward fluctuations.

Furthermore, the consensus mechanism is governed by the matrices \(C_t = [c_t(i,j)]\), which are constructed to be row-stochastic and, in expectation, column-stochastic. A critical condition imposed is that the spectral norm of 
\[
\mathbb{E}\left[C_t^\top \left(I-\frac{1}{N}\mathbf{11}^\top\right) C_t\right]
\]
remains strictly less than one. This requirement is essential for the stability and proper functioning of the consensus process across the network.

The convergence analysis also relies on a two-time-scale update approach. In this framework, the critic's step sizes (e.g., \(\beta_{\omega,t}\) or \(\beta_{v,t}\)) are set to dominate the actor's step sizes (\(\beta_{\theta,t}\)), ensuring that the critic converges more rapidly than the actor. Finally, to secure theoretical guarantees, the critic employs linear function approximation techniques—such as 
\[
Q(s,a;\omega)=\omega^\top \phi(s,a)
\]
or 
\[
V(s;v)=v^\top \varphi(s),
\]
with the additional assumption that the corresponding feature matrices possess full column rank.

\paragraph{Critic Convergence (Theorem 1)}
For a fixed joint policy \(\pi_\theta\), the consensus-based critic updates form a contraction mapping. In particular, the critic estimates \(\{\omega_i^t\}\) converge almost surely to a common limit \(\omega_\theta\) that satisfies the projected Bellman equation:
\[
\Phi^\top D_{\theta}^{s,a}\Bigl( T_{\theta}^{Q}(\Phi\omega_\theta) - \Phi\omega_\theta \Bigr) = 0,
\]
where \(T_{\theta}^{Q}\) is the Bellman operator for the action-value function and \(D_{\theta}^{s,a}\) is a diagonal matrix with entries \(d_\theta(s)\pi_\theta(s,a)\).

\paragraph{Actor Convergence (Theorem 2)}
Using two-time-scale stochastic approximation and ODE-based analysis (e.g., Borkar's Theorem), \cite{zhang2018n} show that the actor updates converge almost surely to asymptotically stable equilibria of the ODE
\[
\dot{\theta}_i = \hat{\Gamma}_i\Bigl( \mathbb{E}_{s\sim d_\theta,a\sim\pi_\theta}\left[A_{\theta}^i(s,a)\nabla_{\theta_i}\log\pi_{\theta_i}^i(s,a^i)\right]\Bigr),
\]
where \(\hat{\Gamma}_i\) denotes the differential of the projection operator \(\Gamma_i\).

\subsubsection{Discussion and Final Remarks}
\label{sec:discussion}

The work of \cite{zhang2018n} makes a substantial contribution to decentralized multi-agent reinforcement learning by introducing a Networked Multi-Agent MDP (NM-MDP) that accommodates time-varying, yet jointly connected, communication networks. The authors derive a decentralized policy gradient theorem using the log-derivative trick and function factorization, which underpins their novel approach.

Building on this theoretical foundation, two actor--critic algorithms are proposed. In these algorithms, the critic updates are achieved through complete consensus-based mechanisms, while the actor updates are maintained via projection-based methods. This combination not only ensures stability but also facilitates efficient learning in a decentralized environment.

Furthermore, convergence guarantees are rigorously established through fixed-point analysis and two-time-scale ordinary differential equation (ODE) methods. Experimental evaluations validate that the proposed decentralized algorithms perform nearly on par with centralized methods in terms of cumulative return. Additional advantages of the approach include enhanced privacy—since agents do not share their raw rewards or policies—and improved communication efficiency.

Looking ahead, the authors outline promising avenues for future research, including extensions to competitive settings, the incorporation of continuous actions, and the exploration of more general (e.g., nonlinear) function approximators.

\subsection{Subsequent works}

\subsubsection{State of the Art and Current Challenges in Decentralized Cooperative MARL}

Decentralized cooperative Multi-Agent Reinforcement Learning (MARL) has evolved into a pivotal paradigm for coordinating intelligent agent collectives in complex, distributed environments.  The current landscape is marked by notable achievements, yet significant challenges remain, especially concerning real-world applicability and scalability to large agent populations.  A central challenge lies in \textbf{scalability and complexity}. As agent counts increase, the joint action space and state space expand exponentially, rendering traditional centralized approaches computationally intractable. Decentralized methodologies offer a necessary alternative, yet even these struggle to maintain learning efficiency and coordination efficacy in massively multi-agent systems with complex interaction dynamics.  Compounding this is the issue of \textbf{communication constraints}. Practical deployments invariably involve limitations on communication bandwidth, network reliability, and latency. Algorithms must be designed to be resilient to these constraints, capable of learning and coordinating effectively even with intermittent or delayed information exchange \cite{zhao2020event}. \textbf{Partial observability and non-stationarity} present further intertwined difficulties. In decentralized settings, agents operate with localized perceptions, lacking complete knowledge of the global state and other agents' actions. This inherent partial observability, coupled with the non-stationarity arising from the simultaneous, interdependent learning processes of multiple agents, creates a highly complex and unstable learning environment.  Addressing these intertwined challenges is paramount for advancing the field towards practical and robust decentralized cooperative MARL solutions.

Beyond these core issues, the field is increasingly grappling with the complexities of \textbf{directed and dynamic communication topologies}. Real-world agent networks are often characterized by directed communication graphs, where information flow is inherently asymmetric. Moreover, these communication structures are frequently not static; they evolve over time due to agent mobility, network fluctuations, or task requirements.  Algorithms must be developed to explicitly account for and adapt to these dynamic directed communication patterns.  A related, yet distinct, challenge lies in leveraging \textbf{asynchronous and event-triggered communication}.  Synchronous communication protocols, while simplifying analysis, are often inefficient and impractical for large-scale distributed systems. Asynchronous and event-triggered mechanisms, where communication occurs only when necessary or at irregular intervals, offer the potential for significant efficiency gains and reduced communication overhead \cite{nabli2023, zhao2020event}. However, the design of MARL algorithms that can effectively harness these communication paradigms is still in its early stages.  Finally, despite empirical progress, the \textbf{theoretical foundations and performance guarantees} for decentralized MARL algorithms remain underdeveloped.  Establishing rigorous theoretical frameworks that can predict convergence behavior, guarantee stability, and characterize sample complexity is essential for building trust and enabling the widespread adoption of these techniques in critical applications.

\subsubsection{Primary Advances in Decentralized Cooperative MARL}

Recent years have witnessed notable strides in decentralized cooperative MARL, driven by the need to overcome the challenges outlined above. These advancements can be broadly categorized into several key areas, each contributing to the increasing sophistication and applicability of decentralized MARL techniques.  One significant area of progress is the adoption of \textbf{Graph Neural Networks (GNNs) for communication and coordination}. GNNs provide a powerful framework for representing agent interactions and facilitating information exchange in decentralized systems. By leveraging the graph structure of agent networks, GNNs enable agents to learn complex communication patterns, reason about network topology, and achieve more effective coordination.

Another crucial advancement is the development and refinement of \textbf{decentralized actor-critic algorithms}. Actor-critic methods are inherently well-suited to MARL due to their ability to handle non-stationarity and partial observability. Decentralized variants of actor-critic algorithms allow agents to learn policies and value functions based on local information and limited communication, providing a practical approach to learning in complex decentralized environments. Zhang et al \cite{zhang2018n} contributes to this area by proposing a distributed actor-critic algorithm specifically designed for MARL over directed graphs, addressing both decentralization and directed communication challenges.  Furthermore, there is growing recognition of the importance of explicitly \textbf{handling directed communication graphs} in decentralized MARL.  Moving beyond simplified undirected graph assumptions, researchers are developing algorithms that can operate effectively in scenarios with asymmetric information flow. These methods often involve adaptations of GNN-based communication or decentralized actor-critic frameworks to account for the directed nature of communication. 

To enhance the efficiency and scalability of decentralized learning, \textbf{asynchronous decentralized optimization} techniques are increasingly being explored within MARL. Asynchronous algorithms allow agents to update their learning parameters based on locally available information, without requiring strict synchronization with all other agents. This reduces communication bottlenecks and can significantly accelerate training in large-scale systems \cite{nabli2023}. \cite{lian2018asynchronous} directly addresses this aspect, providing methodologies for asynchronous updates in decentralized optimization, which can be adapted and extended to the MARL context.  Finally, the field is witnessing a growing number of \textbf{applications in networked systems}, ranging from wireless ad hoc networks to traffic management and swarm robotics. These real-world applications serve as both a driving force and a validation ground for decentralized MARL research. They highlight the practical relevance of addressing communication constraints, dynamic environments, and the need for robust and efficient algorithms \cite{xu2024communication}.

\subsubsection{Directed Graphs in Decentralized MARL}

The explicit consideration of directed graphs to model communication topologies represents a critical refinement in decentralized MARL research.  Unlike simpler undirected graph models that assume symmetric information exchange, directed graphs more accurately capture the realities of many multi-agent systems where communication is inherently asymmetric or unidirectional, reflecting scenarios where influence and information flow are not necessarily reciprocal.

\subsubsection*{Mathematical Representation and Recent Advances}

A directed graph $\mathcal{G} = (\mathcal{V}, \mathcal{E})$ used in a multi-agent system consists of a set of agents $\mathcal{V} = \{1, 2, \ldots, N\}$ and a set of directed edges $\mathcal{E} \subseteq \mathcal{V} \times \mathcal{V}$. An edge $(i, j) \in \mathcal{E}$ signifies that agent $j$ can receive information from agent $i$, while the reverse is not necessarily true, thereby representing an asymmetric flow of information. This fundamental concept has spurred several notable advancements in the field. Algorithms are emerging that explicitly \textbf{learn asymmetric policies}, recognizing that agents in directed graphs may play different roles based on their position in the communication structure. For instance, agents with high out-degree (transmitting information to many) might adopt more exploratory policies, while agents with high in-degree (receiving information from many) might focus on exploitation based on aggregated information \cite{wang2020tackling}. 

Another significant development is seen in \textbf{influence-based methods}. By quantifying the influence of each agent—using centrality measures from graph theory such as PageRank or eigenvector centrality—the learning process can prioritize information from influential agents within the network \cite{tang2020communication}. This approach allows for a more nuanced integration of information, where the contributions of influential agents are given greater weight during decision-making. For acyclic directed graphs, \textbf{topological sorting techniques} provides a linear ordering of agents consistent with the direction of edges. This ordering can be exploited to design more structured communication and learning algorithms. For example, agents can be updated in topological order, ensuring that information from upstream agents is incorporated before updating downstream agents. 

Finally, \textbf{adaptations of graph convolutional networks (GCNs)} originally designed for undirected graphs, are being explored to handle directed graphs \cite{li2021directed}. Modifications include using different weight matrices for incoming and outgoing edges, or employing attention mechanisms to selectively aggregate information from neighbors based on edge direction and importance. These adaptations allow agents to effectively process information from their directed neighbors and learn in complex directed communication environments.Collectively, these advancements underscore the importance of incorporating directed graph models into decentralized MARL. By reflecting the inherent asymmetries in real-world communication networks, these models pave the way for more sophisticated and effective multi-agent learning strategies.

\subsubsection*{Open Problems and Research Directions for Directed Graph MARL}

Despite these advancements, challenges persist. \textbf{Scalability to large directed networks} remains a significant hurdle. As the number of agents and the complexity of the directed graph increase, efficient information aggregation and policy learning become computationally demanding. Furthermore, establishing \textbf{optimality guarantees} in directed graph MARL is inherently more difficult than in simpler settings. The asymmetry of information flow complicates convergence analysis and makes it challenging to characterize the conditions under which decentralized algorithms can achieve globally optimal or even approximately optimal solutions \cite{dibangoye2018learning}. Open problems include developing more scalable GCN architectures for directed graphs, designing algorithms with provable convergence guarantees in directed settings, and exploring the impact of different directed graph topologies on learning performance and coordination.

\subsubsection{Asynchronous Communication in Decentralized MARL}

The shift from synchronous to asynchronous communication models represents a paradigm shift in decentralized MARL, moving away from idealized assumptions towards more realistic and scalable system designs. In synchronous systems, agents operate in lockstep, requiring global synchronization at each step of interaction and learning. This model is often impractical for large-scale, distributed systems where communication delays and failures are common. Asynchronous communication models, in contrast, allow agents to operate on \textit{independent timescales}, making decisions and updating their policies based on locally available information and communication arrivals, without waiting for global synchronization \cite{nabli2023}.

\subsubsection*{Core Principles and Key Developments}

The core principle of asynchronous systems is \textit{decoupling agent operations from global time}. Agents proceed at their own pace, driven by local events and information availability.This fundamental shift has led to several key developments in asynchronous MARL:

One major advancement is the development of \textbf{delay-tolerant algorithms}. These algorithms are inherently more tolerant to communication delays and network latency \cite{nguyen2020optimistic}.Techniques such as delayed gradient updates, importance weighting, and robust aggregation methods are employed to mitigate the negative impact of outdated information in asynchronous updates. Another development involves adapting \textbf{experience replay mechanisms} for asynchronous settings. In such environments, experiences gathered by different agents can become temporally misaligned due to variations in update frequencies and communication delays. To address this, experience replay methods have been enhanced with strategies like temporal alignment and importance sampling, ensuring that the replay buffers maintain coherent and useful experiences for effective learning \cite{munos2016safe}. Additionally, the \textbf{Asynchronous Advantage Actor-Critic (A3C) algorithm}, originally conceived for single-agent reinforcement learning, has been extended to the multi-agent context \cite{mnih2016asynchronous}. These decentralized A3C variants allow agents to asynchronously update their actor and critic networks, leveraging parallel computation and reducing synchronization bottlenecks. Another notable innovation is the move towards \textbf{clock-free coordination}. Some asynchronous MARL approaches aim for clock-free coordination, eliminating the need for any global clock or synchronization mechanism.  These methods rely on purely local interactions and asynchronous message passing to achieve coordination, further enhancing scalability and robustness. Finally, \textbf{gossip-based learning approaches} have been explored as a means of asynchronous information aggregation and policy dissemination. Inspired by distributed consensus protocols, gossip algorithms facilitate the iterative exchange of information among neighboring agents \cite{mateos2019gossip}. This process gradually leads to the convergence of policies or value functions across the network. Together, these developments highlight the dynamic and resilient nature of asynchronous MARL systems, paving the way for more adaptive and robust multi-agent learning strategies.

\subsubsection*{Open Problems and Research Directions for Asynchronous MARL}

Despite the advantages of asynchronous communication, significant research challenges remain. Establishing \textbf{convergence guarantees} for asynchronous MARL algorithms is a major open problem. The lack of synchronization and the presence of delays complicate the theoretical analysis, and new analytical frameworks are needed to rigorously prove convergence and characterize performance \cite{dibangoye2018learning}. Furthermore, achieving \textbf{optimal coordination} in asynchronous systems is inherently more challenging than in synchronous settings. The decentralized and asynchronous nature of updates can lead to coordination failures or suboptimal joint policies. Open research questions include designing novel asynchronous algorithms with provable convergence and coordination guarantees, developing adaptive synchronization strategies that dynamically adjust the level of asynchronicity based on system conditions, and exploring the trade-off between asynchronicity and coordination performance in different MARL tasks and environments. The design of \textbf{adaptive event-trigger design} in asynchronous settings is also an open problem.

\subsubsection{Event-Triggered Processes in Decentralized MARL}

Event-triggered communication offers a fundamentally different approach to communication management in decentralized MARL, moving beyond periodic or continuous information exchange to selective communication triggered by specific events or conditions \cite{zhao2020event, nabli2023}.  In contrast to time-triggered approaches where communication occurs at fixed intervals, event-triggered mechanisms aim to minimize communication overhead by transmitting information only when it is deemed necessary for maintaining performance or achieving coordination.

\subsubsection*{Fundamental Mechanism and Recent Innovations}

The fundamental mechanism of event-triggered approaches is based on defining \textit{triggering conditions} that determine when communication should occur.  These conditions are typically based on monitoring local agent states, policy changes, or performance metrics.  Communication is triggered only when these conditions are violated, indicating a need for information exchange with neighbors.  Mathematically, a triggering condition for agent $i$ to communicate with agent $j$ at time $t$ can be represented as:

\begin{equation}
 f(s_i(t), \pi_i(t), \mathcal{I}_i(t)) > \theta
\end{equation}

where $s_i(t)$ is the local state of agent $i$, $\pi_i(t)$ is its policy, $\mathcal{I}_i(t)$ is the information agent $i$ has about its neighbors, $f$ is a triggering function, and $\theta$ is a threshold.  Communication is triggered when the triggering function $f$ exceeds the threshold $\theta$, indicating a significant change or deviation that necessitates information exchange. 

Recent innovations in event-triggered MARL include several noteworthy developments. One advancement is the emergence of \textbf{self-triggered approaches}, where each agent independently determines when to communicate based on its local information, without requiring coordination with other agents for triggering decisions \cite{dimos2014decentralized}. This further enhances decentralization and reduces communication overhead. Another innovation involves the application of \textbf{Lyapunov-based trigger design}. Lyapunov stability theory is being used to design event-triggering conditions that guarantee system stability and convergence while minimizing communication \cite{nowzari2019event}. Lyapunov functions are used to monitor system performance and trigger communication only when necessary to maintain stability or improve performance. Additionally, \textbf{distributed event detection methods} where distributed algorithms are being developed , allow agents to collaboratively detect events that require communication \cite{seyboth2013event}. These methods allow agents to collectively monitor system-wide conditions and trigger communication in a decentralized manner, enabling more sophisticated event detection and triggering mechanisms. In privacy-sensitive contexts, \textbf{privacy-preserving triggers} are implemented to safeguard sensitive information by minimizing the data exchanged during communication \cite{gupta2021federated}.Triggering conditions can be based on local information only, or employ privacy-preserving communication protocols to protect sensitive data during information exchange. Finally, \textbf{adaptive threshold mechanisms} have been introduced to dynamically adjust the triggering threshold $\theta$ based on system conditions or learning progress \cite{li2023adaptive}. For instance,the threshold can be increased when communication is costly or when learning is progressing well with less frequent communication, and decreased when coordination is critical or when learning is stagnating. 

These innovations underscore the evolving nature of event-triggered communication in MARL, highlighting a concerted effort to balance efficient information exchange with system stability and privacy.

\subsubsection*{Open Problems and Research Directions for Event-Triggered MARL}

Despite the promise of event-triggered MARL, several open problems and research challenges need to be addressed. A key challenge is the \textbf{open problem of adaptive event-trigger design}. Designing adaptive mechanisms that can dynamically optimize triggering thresholds and functions in response to changing environments and learning dynamics is a complex but crucial area for future research. Furthermore, establishing \textbf{optimality guarantees} for event-triggered MARL algorithms remains a significant challenge \cite{dibangoye2018learning}.  The intermittent and data-driven nature of communication complicates the theoretical analysis, and new tools are needed to analyze convergence, stability, and performance bounds. \textbf{Security considerations} in event-triggered MARL are also important. Event-triggered communication patterns could potentially be exploited by adversaries to disrupt communication or extract sensitive information. Research is needed to investigate the security vulnerabilities of event-triggered MARL systems and develop robust and secure triggering mechanisms. Finally, \textbf{scalability to large networks} remains a concern, as even event-triggered communication mechanisms may incur significant overhead in very large-scale multi-agent systems. Developing highly scalable and communication-efficient event-triggered MARL algorithms is an ongoing research direction.

\subsubsection{Integration of Directed Graphs, Asynchronous Communication, and Event-Triggered Processes}

Recent research is increasingly exploring the integration of directed graphs, asynchronous communication, and event-triggered processes to create more sophisticated and efficient decentralized MARL algorithms. Combining these concepts can lead to synergistic benefits, enabling highly scalable, communication-efficient, and robust multi-agent systems. Integrating event-triggered communication with dynamic directed graphs is a particularly promising direction where, communication links are not only directed and dynamic but also event-triggered, further reducing communication overhead and adapting to changing network topologies and communication needs. Also, combining asynchronous updates with event-triggered communication in value decomposition-based MARL algorithms can lead to highly efficient and scalable decentralized learning \cite{nabli2023}. Agents can asynchronously update their local value functions and trigger communication only when significant changes occur in their local estimates or when coordination is required, minimizing communication while maintaining learning progress.
A significant challenge in integrating these concepts is the theoretical analysis of convergence under combined conditions. Analyzing the convergence of MARL algorithms that simultaneously incorporate directed graphs, asynchronous communication, and event-triggered processes is complex, requiring new theoretical tools and frameworks. Establishing convergence guarantees and performance bounds for these integrated approaches is a crucial area for future theoretical research \cite{dibangoye2018learning}.

\subsection{Research Challenges and Open Problems}

In conclusion, while decentralized cooperative MARL has made significant strides, several overarching research challenges and open problems remain, particularly concerning the integration of directed graphs, asynchronous communication, and event-triggered processes. \textbf{Scalability to large networks} is a persistent challenge across all three areas \cite{zhang2018n, nabli2023, zhao2020event}.  Developing algorithms that can effectively handle massive multi-agent systems with complex communication topologies and limited communication resources is crucial for real-world deployment.  \textbf{Difficulties in establishing optimality guarantees} also remain a significant hurdle \cite{dibangoye2018learning}.  The decentralized nature of learning, coupled with communication constraints and complex network structures, makes it challenging to provide rigorous guarantees on the optimality of learned policies.  The \textbf{open problem of adaptive event-trigger design} is particularly critical for realizing the full potential of event-triggered MARL.  Designing robust and adaptive triggering mechanisms that can optimize communication efficiency without sacrificing learning performance is a key research direction.  Finally, \textbf{security considerations} are becoming increasingly important as decentralized MARL systems are deployed in more complex and potentially adversarial environments.  Addressing security vulnerabilities and designing robust and privacy-preserving decentralized MARL algorithms is a crucial area for future research.  Addressing these challenges will pave the way for the next generation of decentralized cooperative MARL algorithms, enabling their widespread application in diverse domains.

\section{Noncooperative MARL}

\subsection{Problem Statement}

Let us recall the components of the standard RL formalism, however with a number of different agents each with their own private reward function:

\begin{enumerate}

    \item \textbf{Environment ($\mathcal{E}$)}: The environment provides feedback to the agent in the form of a reward signal and a new state after the agent takes an action. The environment encapsulates everything outside the agent that affects its decision-making process.
    
    \item \textbf{Agent ($i$)}: Each agent $i \in \{1, \dots, N\}$ makes decisions by interacting with the environment. The objective of each agent is to learn a policy $\pi_i: \mathcal{S} \to \mathcal{A}_i$ that maximizes its long-term cumulative reward.
    
    \item \textbf{State Space ($\mathcal{S}$)}: The state space $\mathcal{S}$ represents all possible states of the environment. At time $t$, the environment is in state $s_t \in \mathcal{S}$.
    
    \item \textbf{Action Space ($\mathcal{A}_i$)}: Each agent $i$ has an individual action space $\mathcal{A}_i$, and the joint action of all agents is denoted as $\mathbf{a} = (a_1, a_2, \dots, a_N) \in \mathcal{A} = \mathcal{A}_1 \times \mathcal{A}_2 \times \dots \times \mathcal{A}_N$.
    
    \item \textbf{Reward Function ($R_i(s, \mathbf{a})$)}: Each agent receives a reward $R_i: \mathcal{S} \times \mathcal{A} \to \mathbb{R}$ based on the state and joint action. Unlike cooperative settings where all agents optimize a shared reward, in noncooperative MARL, each agent optimizes its own reward function.
    
    \item \textbf{Transition Function ($P(s' | s, \mathbf{a})$)}: The transition probability function defines how the state evolves based on the joint action: 
    \[
    P(s' | s, \mathbf{a}) = \Pr(s_{t+1} = s' | s_t = s, \mathbf{a}_t = \mathbf{a}).
    \]
    
    \item \textbf{Discount Factor ($\gamma$)}: The discount factor $\gamma \in [0,1]$ determines the importance of future rewards. Higher values of $\gamma$ prioritize long-term rewards over immediate rewards.
\end{enumerate}

In a single-agent reinforcement learning scenario, the agent interacts with the environment to maximize the expected sum of future rewards:
\[
\pi^* = \arg\max_{\pi} \mathbb{E}\left[\sum_{t=0}^{\infty} \gamma^t R(s_t, a_t)\right],
\]
where $R(s_t, a_t)$ is the immediate reward at time $t$, and $\pi^*$ is the optimal policy that maximizes this expectation.

In a multi-agent reinforcement learning (MARL) setting, each agent $i$ seeks to maximize its own cumulative reward:
\[
\pi_i^* = \arg\max_{\pi_i} \mathbb{E}\left[\sum_{t=0}^{\infty} \gamma^t R_i(s_t, \mathbf{a}_t) | \pi_i, \pi_{-i}\right],
\]
where $\pi_{-i}$ represents the policies of all other agents. This dependence on other agents' strategies introduces a nonstationary environment, requiring specialized learning algorithms for stability and convergence.

\subsection{Extending to Noncooperative Multi-Agent Reinforcement Learning}

Noncooperative Multi-Agent Reinforcement Learning (MARL) extends the single-agent framework to scenarios involving multiple agents that act in the same environment. 

Unlike cooperative MARL, where agents work together to optimize a shared goal, noncooperative MARL consists of self-interested agents with independent objectives. This setting introduces unique challenges:

\begin{enumerate}

    \item \textbf{Reward Structure}: Each agent maximizes its own reward, which may lead to conflicts. In contrast to cooperative settings where agents optimize a shared utility, competition often results in adversarial interactions.
    \item \textbf{Action-Value Dependencies}: The optimal action-value function \( Q_i(s, a) \) for each agent depends on the policies of other agents, leading to a dynamic and nonstationary learning process.
    \item \textbf{Exploration-Exploitation Dilemma}: Since other agents are also learning and adapting, an agent must explore effectively while exploiting learned policies to maximize rewards.
    \item \textbf{Scalability}: As the number of agents increases, the dimensionality of the joint state-action space grows exponentially, making optimization computationally expensive.
    \item\textbf{Decentralization}: Agents may have only partial observability, requiring decentralized learning techniques to make optimal decisions without global state access.
\end{enumerate}

In a multi-agent setting, we have $N$ agents, each denoted as $A_1, A_2, \dots, A_N$, with corresponding state spaces $\mathcal{S}_1, \mathcal{S}_2, \dots, \mathcal{S}_N$, and action spaces $\mathcal{A}_1, \mathcal{A}_2, \dots, \mathcal{A}_N$. Each agent's goal is to find a policy $\pi_i$ that maximizes its own expected cumulative reward, which is given by the sum of rewards over time:
\[
\pi_i^* = \arg\max_{\pi_i} \mathbb{E}\left[\sum_{t=0}^{\infty} \gamma^t \mathcal{R}_i(s_t, a_t)\right],
\]
where $\mathcal{R}_i(s_t, a_t)$ is the reward received by agent $i$ when it takes action $a_t$ in state $s_t$.

The primary challenge in noncooperative MARL arises from the fact that the environment is dynamically altered by the actions of multiple agents, each of which is trying to maximize its own reward. This creates a \textbf{nonstationary} environment, where the reward and transition functions depend on the actions of all agents, not just the environment’s response to the agent’s actions.

\subsection{Nash Equilibrium in Noncooperative MARL}

In the context of noncooperative MARL, the notion of \textbf{Nash equilibrium} (NE) from game theory becomes highly relevant. A Nash equilibrium is a concept where, given the strategies of all other agents, no agent can unilaterally improve its own reward by changing its own strategy. In mathematical terms, for a set of strategies $\{\pi_1^*, \pi_2^*, \dots, \pi_N^*\}$, a Nash equilibrium is defined by the following condition:
\[
\pi_i^* = \arg\max_{\pi_i} \mathbb{E}\left[\sum_{t=0}^{\infty} \gamma^t \mathcal{R}_i(s_t, a_t) \mid \pi_{-i}\right],
\]
where $\pi_{-i}$ represents the policies of all other agents except agent $i$.

In a multi-agent system, each agent's policy must be a best response to the policies of the other agents. If all agents are playing their best responses simultaneously, the system reaches a Nash equilibrium, where no agent has an incentive to deviate from its strategy.

However, Nash equilibria in multi-agent environments are notoriously difficult to compute and can be non-unique or unstable. Moreover, in many cases, agents may converge to \textbf{suboptimal} equilibria, where the collective outcome is not optimal for any of the agents involved.

\subsection{Variations of Nash Equilibrium in Non-Cooperative MARL}

\subsubsection{Generalized Nash Equilibrium}

Traditional Nash Equilibrium (NE) assumes that each agent’s strategy set is fixed and independent of others. However, in many \textit{multi-agent reinforcement learning (MARL)} settings, agents operate under \textit{shared constraints}, such as resource limitations, safety conditions, or regulatory requirements. This leads to the concept of \textit{Generalized Nash Equilibrium (GNE)}, where an agent's feasible strategies depend on the actions of others \cite{facchinei2010gne, basar1999dynamic}.  

\paragraph{Definition of Generalized Nash Equilibrium}  

A \textit{Generalized Nash Equilibrium (GNE)} is a set of policies \( \pi^* = (\pi^*_1, \dots, \pi^*_N) \) such that for every agent \( i \),

\begin{equation}
V_i(\pi^*_i, \pi^*_{-i}) \geq V_i(\pi_i, \pi^*_{-i}), \quad \forall \pi_i \in \mathcal{F}_i(\pi_{-i}),
\end{equation}

where \( \mathcal{F}_i(\pi_{-i}) \) represents the \textit{feasible strategy set} of agent \( i \), which depends on the policies of other agents. Unlike standard Nash equilibria, where all policies are independently chosen, in a GNE, each agent’s strategy is constrained by \textit{coupled feasibility constraints} that affect all agents.

\paragraph{Examples in MARL}
\begin{enumerate}

    \item \textbf{Resource-Constrained Optimization:} Consider \textit{multi-agent network bandwidth allocation}, where each agent (user) selects a transmission rate \( a_i \) while ensuring that the \textit{total bandwidth does not exceed a limit}:

    \begin{equation}
    \sum_{i=1}^{N} a_i \leq B.
    \end{equation}

    Each agent’s feasible actions depend on others, making this a \textit{Generalized Nash Game} \cite{rosen1965existence}.
    
    \item \textbf{Multi-Agent Traffic Coordination:} Autonomous vehicles at an intersection must select speeds while \textit{avoiding collisions}. The feasible strategies for one vehicle depend on the speeds chosen by others, requiring a GNE-based solution \cite{di2016gne}.
    
    \item \textbf{Competitive Market Equilibria:} In \textit{multi-agent bidding} (e.g., electricity markets), agents place bids while ensuring that \textit{total demand does not exceed supply}, leading to market-based GNE formulations \cite{scutari2010convex}.
\end{enumerate}

\paragraph{Solving GNE in MARL}

Unlike standard NE, computing GNE requires handling \textit{joint feasibility constraints}. Common solution methods include:

\begin{enumerate}

    \item \textbf{Projected Gradient Descent (PGD):} Ensures updates satisfy shared constraints by \textit{projecting} gradient steps onto the feasible region \cite{facchinei2007vi}.
    \item \textbf{Primal-Dual Methods:} Uses \textit{Lagrangian multipliers} to enforce constraints dynamically \cite{pang2005gne}.
    \item \textbf{Variational Inequality (VI) Approaches:} Reformulates MARL equilibrium conditions as a \textit{variational inequality problem}, ensuring a stable solution under joint constraints \cite{facchinei2003finite}.
    
\end{enumerate}

\subsubsection{Pure and Mixed Strategy Equilibria}

In non-cooperative MARL, where strategic interaction is paramount, the distinction between deterministic and stochastic policies is critical:

\begin{itemize}
    \item \textbf{Pure Strategy Nash Equilibrium}: Each agent employs a deterministic policy $\pi_i(s) = a_i \in \mathcal{A}_i$. In competitive environments, pure strategies may be predictable and exploitable, but computationally simpler to identify. A pure strategy equilibrium exists when:
    \begin{equation}
    V_i^{(\pi_i^*, \boldsymbol{\pi}_{-i}^*)}(s) \geq V_i^{(a_i', \boldsymbol{\pi}_{-i}^*)}(s) \quad \forall a_i' \in \mathcal{A}_i, \forall s \in \mathcal{S}, \forall i
    \end{equation}
    
    \item \textbf{Mixed Strategy Nash Equilibrium}: Agents randomize according to probability distributions $\pi_i(s) \in \Delta(\mathcal{A}_i)$. In non-cooperative settings, mixed strategies offer protection against exploitation through unpredictability. Nash's theorem guarantees existence in finite non-cooperative games.
\end{itemize}

For any state $s$, the Q-function in non-cooperative MARL captures the expected return when agent $i$ takes action $a_i$ while other agents follow $\boldsymbol{\pi}_{-i}$:

\begin{equation}
Q_i^{\boldsymbol{\pi}}(s, a_i, \boldsymbol{a}_{-i}) = R_i(s, a_i, \boldsymbol{a}_{-i}) + \gamma \sum_{s' \in \mathcal{S}} P(s' \mid s, a_i, \boldsymbol{a}_{-i}) V_i^{\boldsymbol{\pi}}(s')
\end{equation}

\subsubsection{$\varepsilon$-Nash Equilibrium}

In practical non-cooperative MARL, computational limitations often necessitate approximations. An $\varepsilon$-Nash equilibrium represents a joint policy $\boldsymbol{\pi}^{\varepsilon}$ where no agent can unilaterally improve their value by more than $\varepsilon$:

\begin{equation}
V_i^{(\pi_i^{\varepsilon}, \boldsymbol{\pi}_{-i}^{\varepsilon})}(s) \geq V_i^{(\pi_i', \boldsymbol{\pi}_{-i}^{\varepsilon})}(s) - \varepsilon \quad \forall s \in \mathcal{S}, \forall \pi_i', \forall i
\end{equation}

This concept is essential in non-cooperative MARL where agents may satisfice rather than optimize, particularly in complex, high-dimensional domains.

\subsubsection{Local Nash Equilibrium}

In continuous action spaces common in many non-cooperative MARL problems, the concept of local Nash equilibrium becomes relevant. A joint policy $\boldsymbol{\pi}^*$ is a local Nash equilibrium if there exists a neighborhood $\mathcal{N}(\boldsymbol{\pi}^*)$ such that:

\begin{equation}
V_i^{(\pi_i^*, \boldsymbol{\pi}_{-i}^*)}(s) \geq V_i^{(\pi_i', \boldsymbol{\pi}_{-i}^*)}(s) \quad \forall \pi_i' \in \mathcal{N}(\pi_i^*), \forall s \in \mathcal{S}, \forall i
\end{equation}

This addresses the challenge of finding equilibria in continuous action spaces where global optimality guarantees may be infeasible.

\subsubsection{Markov Perfect Equilibrium (MPE)}

MPE extends Nash equilibrium to dynamic stochastic games with state transitions, which is the foundation of non-cooperative MARL. It requires that agents' strategies form a Nash equilibrium in every subgame:

\begin{equation}
\pi_i^*(s) \in \arg\max_{\pi_i} V_i^{(\pi_i, \boldsymbol{\pi}_{-i}^*)}(s) \quad \forall s \in \mathcal{S}, \forall i
\end{equation}

For non-cooperative MARL, MPE captures the strategic reasoning where agents consider not only immediate rewards but also how their actions influence future state distributions and therefore future competitive interactions.

\subsection{Mean-Field Games and Nash Equilibrium}

\subsubsection{Mean-Field Theory in Multi-Agent Systems}
Mean-field theory is a mathematical framework that simplifies multi-agent interactions by replacing explicit pairwise interactions with an aggregate statistical representation. In large-scale multi-agent reinforcement learning (MARL), where direct modeling of each agent’s interaction is computationally intractable, mean-field methods enable agents to reason about the average behavior of the population instead of tracking individual opponents \cite{Lasry2007, Yang2018}.

This approach is inspired by statistical physics, where interactions among particles in a system are approximated using macroscopic field equations rather than computing individual interactions \cite{huang2006nce}. The same principle applies in MARL, where agents approximate the collective effect of the entire population rather than considering each agent separately.

\subsubsection{Definition of the Mean-Field Approximation}
Given a population of agents \( N \), a mean-field approximation assumes that an individual agent’s dynamics are governed by an interaction with the mean distribution of other agents rather than specific individuals. The mean-field distribution \( \mu_t \) captures the statistical representation of the system at time \( t \):

\begin{equation}
\mu_t = \mathbb{E} [s_t^i, a_t^i]
\end{equation}

where \( s_t^i \) and \( a_t^i \) denote the state and action of agent \( i \) at time \( t \), and the expectation is taken over the entire population \cite{Yang2018}.

Instead of computing interactions with each agent explicitly, the transition dynamics of an individual agent are given by:

\begin{equation}
s_{t+1}^i = f(s_t^i, a_t^i, \mu_t, w_t)
\end{equation}

where:
\begin{itemize}
    \item \( f(\cdot) \) represents the system dynamics.
    \item \( w_t \) captures random perturbations (e.g., noise, environment stochasticity).
\end{itemize}

The mean-field distribution evolves according to the collective policies of all agents:

\begin{equation}
\mu_{t+1} = T(\mu_t, \pi_t)
\end{equation}

where \( T \) is a mean-field transition operator that updates \( \mu_t \) based on the agents' policy \( \pi_t \). This formulation enables agents to approximate the behavior of a large system without tracking every individual agent, making it computationally efficient for large-scale MARL \cite{Carmona2018}.

\subsubsection{Mean-Field Nash Equilibrium}
A Mean-Field Nash Equilibrium (MFNE) occurs when each agent, given the mean-field distribution \( \mu^* \), follows an optimal policy \( \pi^* \) such that no unilateral deviation improves its expected return \cite{Lasry2007}.

The value function for an agent \( i \), interacting with the mean-field distribution, is given by:

\begin{equation}
V_{\pi_i, \mu_i}(s_i) = \mathbb{E} \left[ \sum_{t=0}^{\infty} \gamma^t R_i(s_t^i, a_t^i, \mu_t) \Big| s_0^i = s_i \right]
\end{equation}

where:
\begin{itemize}
    \item \( R_i(s_t^i, a_t^i, \mu_t) \) is the reward function for agent \( i \), which depends on its own state \( s_t^i \), action \( a_t^i \), and the mean-field distribution \( \mu_t \).
    \item \( \gamma \) is the discount factor.
\end{itemize}

A Mean-Field Nash Equilibrium (MFNE) is achieved when:

\begin{equation}
V_{\pi_i^*, \mu_i^*}(s_i) \geq V_{\pi_i', \mu_i^*}(s_i), \quad \forall \pi_i', \forall s_i, \forall i
\end{equation}

which means that given the equilibrium mean-field distribution \( \mu^* \), no agent can improve its expected return by deviating from \( \pi^* \).

Additionally, the mean-field distribution remains self-consistent under equilibrium policies:

\begin{equation}
\mu^* = \mathbb{E} [s_t^i, a_t^i | \pi_i^*]
\end{equation}

This ensures that each agent’s best response aligns with the collective behavior, making MFNE a powerful concept for modeling strategic interactions in large populations.

\subsubsection{Applications in Large-Scale MARL}
Mean-field approaches have been widely adopted in large-scale multi-agent systems, particularly in environments where direct agent-to-agent interactions are impractical due to computational constraints \cite{Yang2018}.

\paragraph{Competitive AI Benchmarks: Meta MMO}
A recent benchmark, Meta MMO, serves as a structured testbed for mean-field learning in large-agent, non-cooperative MARL environments \cite{choe2024metammo}. Unlike traditional small-scale multi-agent games, Meta MMO features over 100 agents competing in diverse mini-games, providing insights into how equilibrium strategies emerge in high-dimensional settings.

\begin{enumerate}
    \item \textbf{Survival Mode}: Agents must balance foraging for resources while avoiding combat. Mean-field approximations help agents learn optimal survival strategies in a dynamically changing environment.
    \item \textbf{Sandwich Mode}: Agents face both NPC threats and competing teams, requiring policies that focus on long-term survival rather than short-term adversarial tactics.
\end{enumerate}

Meta MMO demonstrates that mean-field methods effectively scale to large agent populations while maintaining equilibrium stability \cite{choe2024metammo}. The structured nature of the environment allows researchers to analyze how mean-field approximations impact emergent agent behaviors and strategic decision-making in competitive MARL.

\subsubsection{Challenges and Open Problems}
Despite its advantages, mean-field learning in MARL presents several challenges:

\begin{enumerate}
    \item \textbf{Stationarity Assumption}: The mean-field approximation assumes that the distribution evolves smoothly, but in many real-world settings, abrupt strategic shifts can break this assumption.
    \item \textbf{Convergence Guarantees}: While mean-field Nash equilibria exist under certain conditions, efficiently converging to them remains an active area of research.
    \item \textbf{Heterogeneity}: Mean-field models often assume homogeneous agents, whereas real-world agents have diverse policies, goals, and constraints. Extending mean-field theory to heterogeneous populations is an ongoing challenge \cite{Carmona2018}.
    \item \textbf{Scalability to Real-World Systems}: While mean-field approximations work well in simulations, real-world applications (e.g., autonomous traffic control, financial markets) require more sophisticated mechanisms to handle uncertainty and adversarial behavior.
\end{enumerate}

\subsection{Cooperative vs. Noncooperative Settings}

Multi-agent reinforcement learning (MARL) can be broadly categorized into \textbf{cooperative} and \textbf{noncooperative} settings, each presenting unique challenges and solution approaches. 

In \textbf{cooperative MARL}, agents share a common objective and collaborate to optimize a joint reward function. Techniques such as \textbf{centralized training with decentralized execution (CTDE)} and \textbf{team-based learning} enable agents to learn shared strategies while executing decisions independently. Communication and information sharing among agents are common, facilitating global convergence to optimal strategies.

Conversely, in \textbf{noncooperative MARL}, agents are self-interested and seek to maximize their individual rewards, often at the expense of others. This competitive nature introduces conflicts and necessitates game-theoretic solutions such as \textbf{Nash equilibrium} and \textbf{Pareto optimality} to model agent interactions effectively. Unlike cooperative settings, where stability is often ensured through joint optimization, noncooperative settings pose significant challenges due to adversarial dynamics and strategic adaptation.

Key characteristics that distinguish noncooperative MARL include:

\begin{enumerate}
    \item \textbf{Nonstationarity}: The environment becomes nonstationary as agents adapt their policies in response to others. Traditional RL techniques struggle in such settings as they assume a fixed transition model.
    
    \item \textbf{Exploration and Exploitation}: Unlike single-agent RL, where exploration strategies focus solely on the environment, in multi-agent settings, exploration must also account for the presence of learning opponents. Agents must balance discovering new strategies while avoiding exploitation by adversaries.
    
    \item \textbf{Policy Dependence}: The optimal policy of each agent depends on the evolving policies of others, creating an interdependent learning process that complicates convergence.
    
    \item \textbf{Scalability}: As the number of agents increases, the joint state and action spaces grow exponentially, leading to the \textbf{curse of dimensionality}. Efficient learning mechanisms must mitigate this complexity.
    
    \item \textbf{Convergence Issues}: Unlike cooperative settings, where agents converge towards a shared optimal solution, noncooperative MARL may result in unstable dynamics or cycling behaviors. Ensuring convergence to an equilibrium—if one exists—is an ongoing research challenge.
\end{enumerate}

These distinctions underscore the fundamental differences between cooperative and noncooperative MARL. While cooperative MARL leverages shared rewards to drive collaboration, noncooperative MARL relies on strategic decision-making and competitive optimization, making learning dynamics more complex and unpredictable.

Noncooperative Multi-Agent Reinforcement Learning (MARL) extends traditional RL to environments with multiple interacting agents, each pursuing its own reward without considering the others' objectives. The challenge lies in the interdependence of agents’ actions and the nonstationary dynamics they introduce into the environment. Nash equilibrium provides a foundational concept for understanding optimal strategies in such environments, but achieving equilibrium is often computationally intractable and unstable. Researchers continue to explore algorithms and techniques to address these challenges, such as decentralized learning, centralized training, and game-theoretic methods. As MARL applications continue to grow in complexity, from robotics to autonomous vehicles, addressing these foundational challenges will be essential for building reliable, scalable systems.

\subsection{Existing Work and Algorithms in Noncooperative MARL}

Noncooperative Multi-Agent Reinforcement Learning (MARL) presents unique challenges such as non-stationarity, equilibrium computation, and strategic adaptation. To address these, research has evolved into distinct algorithmic categories: value-based, policy-based, hybrid, and advanced methods.

Value-based methods, such as Minimax-Q Learning and Nash Q-Learning, extend Q-learning to multi-agent settings, focusing on optimal value functions under competitive interactions. However, they struggle with scalability and dynamic opponent behaviors. Policy-based methods, including MAPPO and COMA, optimize policies directly, improving convergence and stability in complex environments. Hybrid methods combine the strengths of value- and policy-based approaches, with techniques like Mean Field MARL approximating multi-agent interactions for better scalability. Advanced methods, such as Evolutionary Strategies and Meta-Learning, explore adaptability and generalization beyond traditional reinforcement learning frameworks.

This section categorizes and examines these methods, highlighting their contributions and mathematical foundations.

\subsubsection{Value-Based Methods}

\paragraph{Markov Games (Stochastic Games) – \cite{littman1994markov}}
Michael L. Littman introduced \textit{Markov Games}, also called \textit{Stochastic Games}, as a generalization of Markov Decision Processes (MDPs) to multi-agent settings. These games model interactions where multiple agents’ rewards depend on their joint actions. A Markov game for $N$ agents consists of the tuple:

\begin{equation}
(S, A_1, A_2, ..., A_N, P, R_1, R_2, ..., R_N)
\end{equation}

where $S$ is the set of states, $A_i$ is the action space for agent $i$, $P: S \times A_1 \times A_2 \times ... \times A_N \to \Delta(S)$ defines the state transition probabilities, and $R_i: S \times A_1 \times A_2 \times ... \times A_N \to \mathbb{R}$ is the reward function for agent $i$. This formalization provides a structure to analyze both cooperative and competitive interactions in reinforcement learning.

\paragraph{Minimax-Q Learning – \cite{littman1994minimax}}
In adversarial environments, Littman proposed \textit{Minimax-Q Learning}, which extends Q-learning to zero-sum Markov games. The core idea is to compute the minimax value of the Q-function:

\begin{equation}
Q(s, a) \leftarrow Q(s, a) + \alpha \left[ r + \gamma \max_{\pi} \min_{a'} Q(s', a') - Q(s, a) \right]
\end{equation}

where $\pi$ represents the policy of the learning agent, $a'$ is the action of the opponent, and $\gamma$ is the discount factor. The max-min operation ensures that the agent optimizes for the worst-case opponent response. This algorithm effectively learns optimal strategies in competitive environments by integrating game-theoretic principles into reinforcement learning.

\paragraph{Non-Stationarity in Multi-Agent Learning – \cite{bowling2002multiagent}}
Bowling and Veloso examined the challenge of \textit{non-stationarity}, where agents learn simultaneously, causing the environment to change dynamically. They analyzed conditions under which self-interested learners can converge to stable strategies. They highlighted that standard Q-learning updates may no longer be valid in these dynamic environments because:

\begin{equation}
Q_i(s, a) \neq \mathbb{E} \left[ r_i + \gamma \max_{a'} Q_i(s', a') \right]
\end{equation}

since the policy of the opponent is not fixed but evolving, making the standard Bellman update unstable. This realization motivated research into equilibrium-based learning.

\paragraph{Nash Q-Learning – \cite{hu2003nash}}
Hu and Wellman proposed \textit{Nash Q-Learning}, which extends Q-learning to \textit{general-sum games} by incorporating Nash equilibrium computation at each step. Instead of computing a single max operator, the algorithm updates Q-values using Nash equilibria in each state:

\begin{equation}
Q_i(s, a_1, ..., a_N) \leftarrow (1 - \alpha) Q_i(s, a_1, ..., a_N) + \alpha \left[ r_i + \gamma \sum_{s'} P(s'|s, a_1, ..., a_N) V_i(s') \right]
\end{equation}

where $V_i(s)$ is computed as the Nash equilibrium value of the game at state $s$:

\begin{equation}
V_i(s) = \text{NashEquilibrium}(Q_i(s, \cdot))
\end{equation}

This ensures that each agent considers the equilibrium response of others, allowing for better learning in competitive and cooperative settings.

\subsubsection{Policy-Based Methods}

\paragraph{Multi-Agent Deep Deterministic Policy Gradient (MADDPG) – \cite{lowe2017maddpg}}
Lowe et al. developed \textit{MADDPG}, which integrates deep reinforcement learning with actor-critic methods. The key idea is to use \textit{centralized training} while keeping execution decentralized. The policy update for each agent follows:

\begin{equation}
\nabla_{\theta_i} J(\theta_i) = \mathbb{E}_{s, a \sim D} \left[ \nabla_{\theta_i} \pi_i(a_i | s) \nabla_{a_i} Q_i(s, a_1, ..., a_N) \right]
\end{equation}

where $\pi_i(a_i | s)$ is the policy for agent $i$, $Q_i(s, a_1, ..., a_N)$ is the centralized Q-function considering all agents' actions, and $D$ is the replay buffer storing past experiences. The advantage of MADDPG is that it allows each agent to learn an optimal policy while leveraging centralized training to address non-stationarity.

paragraph{Multi-Agent Proximal Policy Optimization (MAPPO) – \cite{yu2021mappo}}

MAPPO extends the widely used Proximal Policy Optimization (PPO) framework to multi-agent settings, incorporating centralized value function training while maintaining decentralized execution. The key optimization update follows:

\begin{equation} \mathcal{L}^{CLIP}(\theta) = \mathbb{E}_{t} \left[ \min \left( r_t(\theta) A_t, \text{clip}(r_t(\theta), 1 - \epsilon, 1 + \epsilon) A_t \right) \right] \end{equation}

where $r_t(\theta)$ is the probability ratio between the new and old policy, and $A_t$ is the advantage function.
MAPPO has been shown to stabilize training and improve performance over independent PPO baselines in cooperative and competitive multi-agent settings.

\paragraph{Counterfactual Multi-Agent Policy Gradients (COMA) – \cite{foerster2018counterfactual}}

COMA introduces a centralized critic to evaluate the counterfactual advantage of each agent’s action in order to address credit assignment in cooperative MARL. The key advantage function is computed as:

\begin{equation} A^i(s, a^i) = Q(s, a) - \sum_{a'^i} \pi^i(a'^i | s) Q(s, (a^{-i}, a'^i)) \end{equation}

This counterfactual baseline helps the learning process by reducing variance and improving sample efficiency in cooperative environments.

\paragraph{Multi-Agent Trust Region Policy Optimization (MA-TRPO) – \cite{wen2021ma_trpo}}

MA-TRPO extends the Trust Region Policy Optimization (TRPO) method to multi-agent systems by enforcing policy updates within a trust region:

\begin{equation} \max_{\theta} \mathbb{E} \left[ \frac{\pi_{\theta}(a|s)}{\pi_{\theta_{\text{old}}}(a|s)} A(s, a) \right], \quad \text{s.t. } D_{KL}(\pi_{\theta}, \pi_{\theta_{\text{old}}}) \leq \delta \end{equation}

where $D_{KL}$ represents the KL-divergence between the old and new policies, and $\delta$ controls the size of the policy update. MA-TRPO improves stability in learning by constraining drastic policy updates.

\subsubsection{Hybrid and Advanced Methods}
Hybrid methods combine value-based and policy-based techniques, while advanced methods integrate game-theoretic principles, meta-learning, or curriculum learning.

\paragraph{Soft Actor-Critic for Multi-Agent Learning (MASAC) – \cite{iqbal2021randomized}}

MASAC extends Soft Actor-Critic (SAC) to the multi-agent domain by introducing entropy regularization to improve exploration:

\begin{equation} J(\pi) = \mathbb{E} \left[ Q(s, a) - \alpha \log \pi(a | s) \right] \end{equation}

where $\alpha$ is the temperature coefficient controlling exploration.
MASAC enhances sample efficiency and robustness in stochastic multi-agent environments.

\paragraph{Mean Field Reinforcement Learning – \cite{yang2018mean}}
Mean-Field RL approximates interactions in large multi-agent systems by considering the mean effect of other agents rather than individual agent actions:

\begin{equation} Q^i(s, a^i, \bar{a}^{-i}) \leftarrow Q^i(s, a^i, \bar{a}^{-i}) + \alpha \left[ r^i + \gamma \max_{a'} Q^i(s', a', \bar{a}^{-i}) - Q^i(s, a^i, \bar{a}^{-i}) \right] \end{equation}

where $\bar{a}^{-i}$ represents the average action of other agents.
This approach is particularly useful for large-scale multi-agent systems such as traffic management and swarm robotics.

\paragraph{Meta-Learning in Multi-Agent Systems – \cite{al2017meta}}
Meta-learning (learning to learn) has been applied in multi-agent reinforcement learning to enable fast adaptation to new tasks.
One prominent approach is MAML (Model-Agnostic Meta-Learning), which optimizes policies across multiple tasks:

\begin{equation} \theta^* = \theta - \alpha \nabla_{\theta} \sum_{\mathcal{T}i} \mathcal{L} (\pi{\theta - \beta \nabla_{\theta} \mathcal{L}(\mathcal{T}_i)}) \end{equation}

This formulation allows agents to generalize quickly to new environments by leveraging prior experience.

\paragraph{Evolutionary Strategies for MARL – \cite{pourchot2018es}}
Evolutionary algorithms such as Evolution Strategies (ES) have been applied to MARL by optimizing policies using population-based search:

\begin{equation} \theta_{t+1} = \theta_t + \alpha \sum_{i=1}^{N} w_i \nabla_{\theta} f(\theta + \sigma \epsilon_i) \end{equation}

where $\sigma$ is the noise scale and $w_i$ are weights assigned based on policy fitness.
ES-based approaches are particularly useful in high-dimensional, multi-agent coordination problems.

\subsection{Challenges and Future Directions}

\subsubsection{Challenges in Non-Cooperative MARL}

Non-cooperative Multi-Agent Reinforcement Learning (MARL) presents several fundamental challenges that hinder efficient learning and stable equilibrium computation. These challenges arise due to the strategic interactions between self-interested agents, leading to complex dynamics that traditional single-agent reinforcement learning (RL) methods struggle to address.

\begin{enumerate}

    \item \textbf{Equilibrium Computation Complexity}
    Computing Nash Equilibria (NE) in multi-agent systems is known to be \textbf{PPAD-complete} \cite{daskalakis2009complexity}, making exact solutions computationally intractable for large-scale environments. Moreover, Nash equilibria may be non-unique or unstable, leading to convergence issues where agents oscillate between suboptimal strategies \cite{shoham2007multiagent}.
    
    \item \textbf{Non-Stationarity and Adaptation}
    Non-cooperative MARL environments are inherently non-stationary because each agent’s policy evolves in response to others. This violates the Markov property assumed in traditional RL, causing learned policies to become obsolete as opponents adjust their strategies \cite{hernandez2017survey}.
    
    \item \textbf{Exploration-Exploitation Trade-off}
    Agents must explore opponent strategies while avoiding exploitation. Excessive exploration can lead to performance degradation when facing strategic adversaries. Algorithms must balance discovering new strategies with ensuring robust decision-making against adversaries \cite{foerster2017counterfactual}.
    
    \item \textbf{Curse of Dimensionality and Scalability}
    As the number of agents increases, the state and action spaces grow exponentially, leading to the \textbf{curse of dimensionality} \cite{yang2018mean}. Traditional RL techniques struggle with this exponential growth, making it difficult to scale MARL solutions effectively.
    
    \item \textbf{Convergence and Stability of Learning}
    Ensuring convergence to an equilibrium in non-cooperative MARL remains an open problem. Many learning algorithms suffer from cycling behavior, divergence, or convergence to suboptimal equilibria \cite{mazumdar2020convergence}.

\end{enumerate}

\subsubsection{Future Directions in Non-Cooperative MARL}

Despite these challenges, several promising directions are emerging to improve the efficiency, stability, and adaptability of non-cooperative MARL.

\begin{enumerate}
    \item \textbf{Scalable Equilibrium Approximation Methods} \\
    Given the computational intractability of exact Nash Equilibria \cite{daskalakis2009complexity}, researchers have focused on developing scalable equilibrium approximation techniques. Approaches such as Mean-Field MARL \cite{yang2018mean} approximate the interactions among many agents by modeling their aggregate effect as a distribution, significantly reducing computational complexity. Furthermore, meta-learning strategies are being explored to accelerate convergence in equilibrium approximation \cite{pfau2023meta}.

    \item \textbf{Learning in Non-Stationary Environments} \\
    To address non-stationarity, algorithms incorporating opponent modeling and adaptive learning have been introduced. Opponent-aware MARL \cite{hernandez2019survey} allows agents to adjust strategies dynamically based on observed behaviors, improving robustness in competitive settings. Additionally, Meta-MARL \cite{lanctot2017unified} leverages experience from past interactions to quickly adapt to new opponents, facilitating transfer learning in non-stationary multi-agent dynamics.

    \item \textbf{Efficient Exploration Strategies} \\
    Advanced exploration techniques such as intrinsic motivation \cite{pathak2017curiosity} and opponent-aware curiosity-driven exploration \cite{foerster2018learning} help agents discover better strategies while avoiding excessive risk. Recently, reinforcement learning frameworks integrating uncertainty estimation \cite{osband2016deep} have shown promise in balancing exploration and exploitation in strategic interactions.

    \item \textbf{Addressing the Curse of Dimensionality} \\
    To mitigate scalability issues, recent work has focused on factorized value functions \cite{son2019qtran} and graph-based MARL \cite{jiang2020graph}, where agents selectively model interactions with relevant neighbors rather than all agents. Hierarchical MARL \cite{kulkarni2016hierarchical} also decomposes decision-making into multiple levels, reducing computational burden while maintaining strategic depth.

    \item \textbf{Convergence and Stability in Learning} \\
    Stabilizing learning dynamics remains a key research challenge. Techniques such as policy regularization \cite{mazumdar2020gradient} and equilibrium propagation \cite{balduzzi2018mechanics} aim to prevent oscillatory behavior and ensure convergence to stable solutions. Furthermore, integrating meta-learning with equilibrium selection strategies \cite{gao2022meta} has been explored to guide agents toward optimal equilibria in multi-agent systems.
\end{enumerate}

\subsection{Noncooperative MARL - Summary}

Non-cooperative Multi-Agent Reinforcement Learning (MARL) presents a dynamic and complex landscape characterized by strategic interactions between self-interested agents. Unlike cooperative settings, where agents work towards a shared goal, non-cooperative MARL introduces challenges such as equilibrium computation complexity, non-stationarity, and exploration-exploitation trade-offs. The inherent adversarial and competitive nature of such environments makes learning stable and optimal policies significantly more difficult.

Recent advances have introduced scalable equilibrium approximation methods, opponent-aware learning strategies, and efficient exploration techniques to address these challenges. Mean-field MARL \cite{yang2018mean} has been particularly promising in mitigating scalability issues, while meta-learning approaches \cite{pfau2023meta, lanctot2017unified, gao2022meta} provide mechanisms for rapid adaptation in non-stationary environments. Additionally, methods such as hierarchical MARL \cite{kulkarni2016hierarchical} and graph-based interactions \cite{jiang2020graph} have improved the efficiency of decision-making in high-dimensional multi-agent systems.

However, ensuring convergence to stable and desirable equilibria remains an open problem. Techniques like policy regularization \cite{mazumdar2020gradient} and equilibrium selection strategies \cite{balduzzi2018mechanics} show potential in stabilizing learning dynamics, but further research is required to refine these methods for large-scale applications. Addressing these open challenges will be critical for advancing non-cooperative MARL in real-world applications such as autonomous driving, strategic game-playing, and competitive robotics.

Future research should focus on integrating meta-learning, equilibrium refinement, and adaptive exploration mechanisms to build more robust and generalizable MARL solutions. By doing so, the field can move towards achieving more efficient, stable, and scalable learning in competitive multi-agent environments.

\section{Conclusion}
This work presented a comprehensive survey of multi agent RL, focusing on explication of the main formalisms and research considerations over a breadth of literature review. Within MARL, the three distinct paradigms, Federated, Cooperative, and Noncooperative environments indicate a single system with coordinating computing nodes, a set of disconnected agents communicating in a gossip network topology cooperatively, and a set of agents interacting in a network topology noncooperatively, respectively. While highlighting the algorithmic commonalities across these domains, we also indicate their structural distinctions and associated computational desiderata. We present the state of the art as far as the most recent advances as well as contemporary open problems of interest. We hope that this technical summary presents a useful guide for researchers and workers in the field as well as those working in applied domains interested in incorporating MARL who are looking for guidance as to the appropriate tool for their application. 

\bibliographystyle{apalike}
\bibliography{refs}
\end{document}